\documentclass[final,5p,times,twocolumn]{elsarticle}

\usepackage{graphicx}
\usepackage{dcolumn}
\usepackage{bm}
\usepackage{amsmath}
\usepackage{amssymb}
\usepackage{mathptmx}
\usepackage{textcomp}
\usepackage{tabularx} 
\usepackage{times}
\usepackage{wrapfig}
\usepackage{caption}
\usepackage{subcaption}
\usepackage{adjustbox}
\usepackage{listings}
\usepackage{silence}
\usepackage{wasysym}
\usepackage{titlesec}
\usepackage{graphicx}
\usepackage{colortbl}
\usepackage[table,xcdraw]{xcolor}
\usepackage{booktabs}  
\usepackage{chemfig}
\usepackage{graphicx}
\usepackage[utf8]{inputenc}
\usepackage[T1]{fontenc}
\usepackage{etoolbox}
\usepackage{csquotes}
\usepackage{xcolor}
\usepackage{appendix}

\usepackage{euscript}   
\usepackage[table]{xcolor} 
\usepackage{hyperref}
\hypersetup{
    colorlinks=true,
    linkcolor=cyan,
    filecolor=magenta,      
    urlcolor=blue}

\definecolor{headergray}{rgb}{0.5, 0.5, 0.5}  

\begin{document}

\title{Spatially Local Estimates of the Thermal Conductivity of Materials}

\date{\today}

\author[LANL]{C. Ugwumadu}
\ead{cugwumadu@lanl.gov}
\affiliation[LANL]{organization = Quantum and Condensed Matter Physics (T-4) Group, addressline = { Los Alamos National Laboratory}, city = {Los Alamos}, postcode = {87545}, state ={NM}, country={USA}}

\author[NQPI]{A. Gautam}
\ead{ag007122@ohio.edu}
\affiliation[NQPI]{organization = Department of Physics \& Astronomy, Nanoscale & Quantum Physics Institute, addressline = {Ohio University}, city = {Athens}, postcode = {45701}, state ={OH}, country={USA}}

\author[INPP]{Y. G. Lee}
\ead{yl518521@ohio.edu}
\affiliation[INPP]{organization = Department of Physics \& Astronomy, Institute of Nuclear & Particle Physics, addressline = {Ohio University}, city = {Athens}, postcode = {45701}, state ={OH}, country={USA}}

\author[NQPI]{D. A. Drabold}
\ead{drabold@ohio.edu}

\begin{abstract}
In this paper we describe a spatial decomposition of the thermal conductivity, what we name "site-projected thermal conductivity", a gauge of the thermal conduction activity at each site. The method is based on the Green-Kubo formula and the harmonic approximation, and requires the force-constant and dynamical matrices and of course the structure of a model sitting at an energy minimum. Throughout the paper, we use high quality models previously tested and compared to many experiments. We discuss the method and underlying approximations for amorphous silicon, carry our detailed analysis for amorphous silicon, then examine an amorphous-crystal silicon interface, and representative carbon materials. We identify the sites and local structures that reduce heat transport, and quantify these (estimate the spatial range) over which these "thermal defects" are effective. Similarities emerge between these filamentary structures in the amorphous silicon network which impact heat transport, electronic structure (the Urbach edge) and electronic transport. 
\end{abstract}


\maketitle

\section{\label{sec:introduction}Introduction}
The transport of heat is a key property of any solid. It is challenging to measure, and it is also difficult to compute and this is a time of lively debate on new theories \cite{AF,FA,Koh,Yang,EMD,NEMD,VCAmethod,Kang,JJ,DiffusonCoherence,simoncelli,Baroni1,SPTC, MLthermal}. From the standpoint of materials design, the ability to predict atomistic or mesoscale transport is valuable. Perhaps the most obvious application of these methods is to thermoelectric materials \cite{Serge}. 

The work we present here is an extension of two classic papers of Allen and Feldman (AF) \cite{AF}, and Feldman, Kluge, Allen and Wooten (FKAW) \cite{FA} in 1993. In these papers, the authors derived the form of the thermal conductivity (TC) in the harmonic approximation for a quantized lattice, and computed the thermal conductivity (TC) for the classic homogeneously disordered material, amorphous silicon (a-Si), using realistic structural models of Wooten-Weaire-Winer type \cite{WWW} and the Stillinger-Weber potential for Si \cite{SW}. These authors offered insights into the nature of heat transport in a-Si, and established a framework to understand the dependence of the transport on the various phonons present in the material. This fundamental work has been developed in many ways \cite{SPTC, Minamitani,Minamitani2,Henry,TC_aSi2, Larkin2014}.

Here, we extend this work to determine site-wise contributions to the TC. The method of AF in its original form is a tool to answer the question: "Given a structure, and force-constant matrix, what is the conductivity tensor"? We show that it is possible to decompose these global quantities into spatially local (site) estimates. We  published a first report of the method [dubbed the "site-projected thermal conductivity (SPTC) \cite{SPTC}].   Beside studying a-Si partly as a check of the codes against the work of AF and FKAW, we applied the method to amorphous graphene, grain boundaries in diamond silicon, and Si-Ge systems to jointly explore mass and chemical disorder on heat transport.  
More generally, we show how to obtain spatially resolved information from a global transport calculation. The Kubo formulae \cite{Kubo} expresses transport coefficients in terms of autocorrelation functions including all the atoms in a cell. Different parts of a disordered network contribute unequally to the transport, and those effects are faithfully represented by the global formula. It is therefore reasonable to attempt to "disentangle" these local contributions. Analogous to the problem of making atomic-level estimates of the electrical conductivity "Space-projected conductivity" \cite{SPC1,SPC2}, it is possible to express the thermal conductivity as a double spatial sum over sites in a cell, and by summing out one of those atomic indexes, a local estimate for TC accrues. 

Our spatial projection or "disentangling" proceeds by expressing the thermal conductivity of AF as a series of nested sums involving the force constant matrix ($FCM$), eigenvalues and vectors of the dynamical matrix ($DM$), positions of the atoms and the lattice vectors for the cell. By reorganizing these sums so that the outermost sums are on the atoms in a cell, we are left with an expression for the thermal conductivity of the form:
\begin{equation}
    \kappa= \sum_{x,x'}\Xi(x,x')
\end{equation}

\noindent where the object $\Xi$ is extracted from the work of AF, and $x$ and $x'$  index sites in the cell. Following the simple logic used in "Mulliken charge population analysis" in quantum chemistry\footnote{Thus, following Szabo and Ostlund \cite{szabo1996, Mulliken}, if $\rho_e$ is the single-particle density matrix and S the overlap matrix in a single-orbital site-centered representation, then if N is the total number of electrons, $N=Tr(\rho_e S)=\sum_{\mu,\nu}{\rho_e}_{\mu \nu}S_{\nu \mu}$ and the charge associated with a particular site $\mu$ is $Q_\mu=(\rho_e S)_{\mu \mu}$.}(in which one estimates a local charge on a given atom in a molecule from the system's density matrix and overlap matrix), we interpret the object $\zeta(x)=\sum_{x'}\Xi(x,x')$ to be a site-projected thermal conductivity. This charge decomposition is not unique, neither is our local estimate for TC, though we find it useful and informative. 

The purpose of the present paper is two fold. First, we provide more details of the formulation of the SPTC and give details  about approximations invoked and convergence of the computations. Secondly, we seek to correlate the form and function of disorder in materials \cite{festshrift} for this case the heat transport. 

The paper is organized as follows. In Section \ref{sec:Theory}, we introduce the SPTC formalism and discussed its computational implementation in Section \ref{sec:Computation}. Next, we  explore its meaning and convergence for a-Si for large and realistic models in Section \ref{sec:Application_aSi}. For the problems involving Si, we use the reliable ML GAP potential \cite{GAP_Si}. We show that it is straightforward to carry out these computations for systems with thousands of atoms. Our results closely reproduce those of the classic papers of AF and FKAW.

In Section \ref{sec:OtherApplication} we examine several applications, including a Si diamond/amorphous interface, Silicon suboxides and sp$^2$ carbon systems. We also observe an interesting correlation between electronic structure/transport and thermal transport. In Section \ref{sec:conclucion} we discuss open questions about the method and is prospects for future development.

\section{\label{sec:Theory}A Local Estimate of Thermal Conductivity}

\subsection{Harmonic Approximation}
Heat transport in materials generally involves both electrons and phonons. In this paper, we restrict our attention to the phonon contribution. By invoking the Harmonic Approximation (HA), phonons emerge as the heat carriers.  The potential energy $U$ is given as:

\begin{subequations}\label{eqn:harmonic}
\begin{gather}
       U = \frac{1}{2}\sum_{\gamma, \gamma'} \sum_{i,j} \sum_{\alpha,\beta}\phi_{ij}^{\alpha \beta}(\gamma, \gamma') u_{i \gamma}^{\alpha} u_{j \gamma'}^{\beta}\\
      \phi_{ij}^{\alpha \beta}(\gamma, \gamma') =  \frac{\partial^2E}{\partial u_{i \gamma}^{\alpha} \partial u_{j \gamma'}^{\beta}}
\end{gather}
\end{subequations}
where $\phi_{i j}^{\alpha \beta}(\gamma, \gamma')$ is the $FCM$,  $\gamma$ and $\gamma'$ represent cells in a periodic system. $ u_{i \gamma}^{\alpha} $ is the $\alpha$ Cartesian component of the displacement vector for $i^{th}$ atom in the $\gamma^{th}$ cell. Using periodicity, we can write $u^{\alpha \mathbf{k}}_{i \gamma} \left( t \right)$ where \textbf{k} is a wave vector that may be chosen within  the Brillouin zone:

\begin{equation}\label{eq:DM_FC}
u^{\alpha \mathbf{k}}_{i \gamma} \left( t \right) = \frac{1}{\sqrt{m_i}} \sum_{m} e^{\alpha m}_{i \mathbf{k}} e^{i \left( \vec{k} \cdot \vec{R}_{i \gamma} - \omega_{m} t \right)},
\end{equation}

where $m_i$ is the mass of the $i^{th}$ atom, $e^{\alpha m}_{i \mathbf{k}} $ is the polarization of the $m^{th}$ vibrational mode (with a total $3 \times N$ modes, for the $N$ number of atoms in the cell) at $i^{th}$ atom along $\alpha$ direction. The classical equation of motion with this \textit{ansatz} leads to the usual eigenvalue problem: 

\begin{subequations}\label{eq:DM_EP}
\begin{align}
{\omega^{\mathbf{k}}_m}^2 e^{\alpha m}_{i \mathbf{k}} &= \sum_{\beta j} D^{\alpha \beta }_{i j} \big( \mathbf{k} \big) e^{\beta m}_{j \mathbf{k}}, \\
D^{\alpha \beta }_{i j}( \mathbf{k} )  &= \sum_{\gamma} \frac{1}{\sqrt{m_i m_j}} \phi_{i j}^{\alpha \beta}(0, \gamma) e^{i \mathbf{k} \cdot (\mathbf{R}_{j \gamma} - \mathbf{R}_{i 0})}
\end{align}
\end{subequations}

where $D^{\alpha \beta }_{i j}(\mathbf{k})$ is the $DM$, the lattice Fourier transform of the $FCM$, $\phi_{i j}^{\alpha \beta}(0, \gamma)$. $\mathbf{R}_{j \gamma}$ is the position of $j^{th}$ atom in the $\gamma^{th}$ cell. The eigenvalues and eigenmodes of $D^{\alpha \beta }_{i j}(\mathbf{k})$ are the vibrational frequency of the $m^{th}$ mode $\omega_m^\mathbf{k}$ and the polarization $e^{\alpha m}_{i \mathbf{k}} $.

\subsection{Cell size, Periodicity and Anharmonicity}
At this point, we assume that our cell is large enough to justify sampling the Brillouin zone only at $\Gamma$. This is done for practical and  fundamental reasons: the thermal (or electrical) conductivity is infinite for any perfect crystal (independent of the number of atoms in the unit cell) \cite{AshcroftMermin}, a point emphasized by Fiorentino \textit{et al.} \cite{Baroni1}. They further discuss the importance of anharmonicity, which is neglected in the present work, but might be included in the Quasi-Harmonic Approximation \cite{Baroni2}. Also,  our calculations are focused on the diffusive regime for periodic systems, in part because the sampling of low energy acoustic-like modes is limited.

We see that $\zeta(x)$,  obtained as the sum over atomic positions in the cell is rapidly convergent for the systems studied in this paper because of the decay of the $FCM$. A  minimum sensible supercell volume to consider for our method is one large enough that there is negligible interaction between a reference atom and its periodic images, and presumably such that the phonon mean free path is significantly smaller than the cell size.   We report results on a-Si for cells ranging from 64 to 4096 atoms. 

\subsection{Site-Projected Thermal Conductivity}
At the $\Gamma$ point of the phonon Brillouin zone,  the vibrational normal modes are real.  For such a case,  the AF expression for TC, represented as $\kappa$, takes the form:

\begin{align} \label{eq:TC_Long}
    \kappa = & \frac{\pi \hbar}{48 T V} \sum_{m, n\neq m} \left[- \frac{\partial \langle f_m\rangle}{\partial \omega_m} \right] \delta(\omega_m - \omega_n ) \frac{\left( \omega_m + \omega_n \right)^2}{\omega_m \omega_n} \notag \\
    & \sum_{\eta} \sum_{\alpha, \beta} \sum_{\gamma, x, x'} e^{\alpha m}_{x} e^{\beta n}_{x'} \frac{1}{\sqrt{m_x m_{x'}}}  \phi^{\alpha \beta}_{x  x'} ({0}, \gamma) \left({R}^{\eta}_{\gamma} + {R}^{\eta}_{x x'}\right) \notag \\
    & \sum_{\alpha', \beta'} \sum_{\gamma', a, b} {e^{\alpha' m}_{a}} {e^{\beta' n}_{b}} \frac{1}{\sqrt{m_a m_b}}  \phi^{\alpha' \beta'}_{a b} ({0}, \gamma') \left({R}^{\eta}_{\gamma '} + {R}^{\eta}_{a b}\right)
\end{align}

where, $m$ and $n$ are the indices of the classical normal modes, $f_m$ is the equilibrium occupation of the $m^{th}$  mode, $\omega_m$ and $e^{\alpha m}_{i} $ are the vibrational frequency $\omega_m^\mathbf{0}$ and the polarization $e^{\alpha m}_{i \mathbf{0}} $ . ${R}_{\gamma}^{\eta}$ is the $\eta^{th}$ component of $\mathbf{R}_{\gamma}$; ${R}_{x x'}^{\eta}$ is the $\eta^{th}$ component of $\mathbf{R}_{x x'}$.  The thermal conductivity in Equation \ref{eq:TC_Long}, is taken as an average of the diagonal components of the conductivity tensor.

Next, we seek to extract local information about thermal conduction within the AF picture. The AF form for TC may be rearranged as a double sum over spatial points (labeled ${x}$).   Carrying this out, with Equation \ref{eq:TC_Long}, we find:
\begin{align}\label{sptc}
    \kappa = & \sum_{x, x'} \Xi(x, x'),
\end{align}
\noindent where
\begin{align}
\label{eq:Gamma_def} 
   \Xi(x, x') &= \frac{\pi \hbar^2}{48k_BT^2V} \frac{1}{\sqrt{m_{x'} m_{x}}} \sum_{\eta} \sum_{\gamma}   \left({R}^{\eta}_{\gamma}  + {R}^{\eta}_{xx'}\right) \sum_{m, n \neq m} \delta(\omega_m - \omega_n) \notag \\
    &  \frac{(\omega_m + \omega_n)^2}{\omega_m \omega_n}  \left( \frac{e^{\frac{\hbar \omega_m}{k_B T}}}{\left(e^{\frac{\hbar \omega_m}{k_B T}} - 1\right)^2}\right)
    \sum_{\alpha \beta} \phi^{\alpha \beta}_{x, x'}(0, \gamma) e^{\alpha m}_{x} e^{\beta n}_{x'}  \notag \\
    &  \sum_{\gamma' a b} \sum_{\alpha' \beta'}\frac{1}{\sqrt{m_a m_b}}  \phi^{\alpha' \beta'}_{a, b} (0, \gamma')  e^{\alpha' m}_{a} e^{\beta' n}_{b} \left(R^{\eta}_{\gamma '} + R^{\eta}_{ab}\right).
\end{align}

$\Xi$ is a real-symmetric matrix with units of thermal conductivity. We decompose the total TC into contributions depending on atomic position $x$ by conduction a summation of $\Xi(x, x')$ over all positions $x '$:
\begin{equation}
\label{eq:SPTC}
\zeta(x) = \sum_{x'} \Xi(x, x').
\end{equation}
We call $ \zeta(x) $ the \textit{site-projected thermal conductivity} (SPTC), interpreted as the contribution of atom at site $x$ in the cell to the total TC of the system, since:
\begin{equation}
\label{eq:kappa_zeta}
    \kappa = \sum_{x} \zeta(x).
\end{equation}
For anisotropic systems or off-diagonal terms the conductivity tensor $\kappa_{\alpha \beta}$ local contributions can be similarly obtained.

\subsection{Spectral Properties of $\Xi$}

The eigenvalue problem for the  (real-symmetric) matrix, $\Xi$ with dimensions of thermal conductivity, reads:
\begin{equation}\label{eqn:EigenValProblem}
    \Xi \eta_\mu=\lambda_\mu \eta_\mu
\end{equation}

 \noindent $\Xi$ is traceless, $\sum_\mu \lambda_\mu =0$,  so that the density of states of $\Xi$ has both positive and negative support. The spectral version (in the $\Xi$ diagonal representation) of the  thermal conductivity is 
\begin{equation}
  \kappa=\sum_{\mu, x,x' (x \neq x')} \lambda_\mu \eta_\mu (x) \eta_\mu (x')  
\end{equation}

\noindent this implies that there is also a "mode-projected" conductivity:
\begin{equation}\label{eqn:kappa_mu}
  \kappa_\mu= \lambda_\mu\sum_{x,x' (x \neq x')}  \eta_\mu (x) \eta_\mu (x')  
\end{equation}

In this representation, the spectral version of the SPTC becomes:

\begin{equation}
  \zeta(x)=\sum_\mu \lambda_\mu[\eta^2_\mu(x)+\sum_{x,x',x \neq x'}\eta_\mu(x) \eta_\mu(x')]
\end{equation}
\noindent We pause for a moment to link this to the electronic transport. The analogous eigenvalue problem for the space-projected electronic conductivity is (\cite{SPC1}):

\begin{equation}
\label{eqn:TheGamma}
    \Gamma \chi_\mu=\Lambda_\mu \chi_\mu
\end{equation}

Here,  $\Gamma$ is obtained from the Kubo-Greenwood formula \cite{Kubo,Greenwood}, as $\Xi$ is from the Green-Kubo formula \cite{Kubo, Green} and the work of AF. $\Gamma$ always displays a huge null space, meaning that the great majority of $\Lambda$ were concentrated near zero. Electronic conduction activity was compactly represented with a small number of ($\Lambda, \chi$) with $\Lambda \neq 0$. 

For both the electronic and thermal cases, the eigenvectors for $\Gamma$ and $\Xi$ may be interpreted as a "modes of conduction" for electrons and heat, respectively. These seem to be akin to the "transmission eigenchannels" of electron transport \cite{EigChannel}, the conjugate eigenvalues being  a transmittance for the particular mode \cite{Landauer1970,Nikoli}. 
We discuss the individual contributions of the $\kappa_\mu$ for the case of a-Si below.

Also, in analogy with electronic structure theory, we can interpret $\Xi$ as a  "thermal density matrix", since in the spectral language of $\Xi$: 
\begin{equation}\label{eqn:Locality}
   \Xi(x,x') =\sum_\mu \lambda_\mu \eta_\mu (x) \eta_\mu (x') =\rho(x,x') 
\end{equation}
from which we can also write:
\begin{equation}
   \kappa=\sum_{x,x'} \rho(x,x')=\sum_{x,x'} \Xi(x,x')  
\end{equation}

Thus the decay of this "thermal density matrix" is determined by the decay of the $FCM$ (see  \ref{AppendixA}).


\section{Computational Implementation}\label{sec:Computation}

The SPTC framework is implemented in the \texttt{C++} programming language using a collection of source files, hereafter referred to as \textit{modules}. These modules adopt a procedural programming approach and are organized into four key stages, each responsible for a distinct part of the thermal transport pipeline. Parallelization is achieved using OpenMP directives \cite{OpenMP}, enabling efficient scaling for large systems with thousands of atoms and dense spectral sampling. The code is optimized for high-performance computation, with binary output files used extensively for storage and subsequent visualization or post-processing.

\subsection{Post-Processing of Interatomic Force Constants}\label{subsec:forceConstant}

The module \textsc{DynMat\_process.cpp} converts the real-space $DM$--typically obtained from first-principles calculations using the Vienna \textit{Ab initio} Simulation Package (\texttt{VASP}) \cite{vasp} or from classical simulations using the Large-scale Atomic/Molecular Massively Parallel Simulator (\texttt{LAMMPS})\cite{lammps}-—into properly mass-weighted harmonic force constants suitable for lattice dynamics and thermal transport analysis. 

The required inputs include atomic species, Cartesian coordinates, atomic masses, system dimensions, and the $DM$ formatted in Cartesian coordinates. The $DM$ is assumed to be a $3N \times 3N$ symmetric matrix representing interatomic force constants for a system of $N$ atoms. These elements are parsed and stored in a 4D array: $[i][j][\alpha][\beta]$, where \(i\) and \(j\) index atoms, and \(\alpha,\beta = 0,1,2\) represent Cartesian directions. Symmetrization is enforced via
\begin{equation}
  \phi_{ij}^{\alpha \beta} \rightarrow \frac{1}{2} \left( \phi_{ij}^{\alpha \beta} + \phi_{ji}^{\beta \alpha} \right)  
\end{equation}

\noindent ensuring physical reciprocity of interatomic interactions. The matrix is then mass-weighted according to:
\begin{equation}
\phi_{ij}^{\alpha \beta} \rightarrow \phi_{ij}^{\alpha \beta} \cdot \sqrt{m_i m_j},
\end{equation}
\noindent where \(m_i\) and \(m_j\) are the atomic masses of atoms \(i\) and \(j\), respectively.

\subsection{Harmonic Approximation and Phonon Mode Extraction}\label{subsec:Harmonic}

The module \textsc{harmonic\_approximation.cpp} performs phonon mode analysis by diagonalizing the mass-weighted $FCM$ in Equation \ref{eqn:harmonic}b. Input data include structural geometry, the harmonic force constants, and the target $\mathbf{k}$-point (typically $\Gamma$). The force constants are remapped into a 7D array: $[i][j][\alpha][\beta][dx][dy][dz]$,
where \texttt{dx}, \texttt{dy}, and \texttt{dz} define the lattice translation vectors mapping atom \(j\) to periodic images with respect to atom \(i\). This array stores the second derivatives of the total energy with respect to atomic displacements as discussed in Equation \ref{eqn:harmonic}a.

From these, the $DM$ diagonalization is performed with LAPACK's \texttt{zheevd} routine \cite{lapack99}, yielding  eigenvectors and  eigenvalues. Outputs include the phonon frequencies (THz), mode-resolved eigenvectors, and corresponding eigenvalues. Atom-resolved eigenmodes are also exported for visualization and projection tasks.

\subsection{Computing the Heat Current}\label{subsec:SPTCpreprocessing}

The \textsc{sptc\_preprocessing.cpp} module constructs the velocity-like coupling tensors required to evaluate site-projected thermal conductivity within the AF framework. It bridges the harmonic mode analysis and the final SPTC computation by producing the mode-resolved couplings \(S_{ij}^{\alpha}(m,n)\) that quantify energy transfer between vibrational modes. We note that \(S_{ij}^{\alpha}(m,n)\) is more commonly known as the heat current operator \cite{Hardy, AF_LPD2}.

The module first reads all necessary data: mass-weighted force constants, phonon eigenvectors and frequencies, atomic positions, atomic masses, and system dimensions. It then constructs neighbor lists using a cutoff radius, and computes the minimum image convention vectors \(\Delta R_{ij}^{\alpha}\) between all atomic pairs.

For each phonon mode pair \((m,n)\), and for each atom pair \((i,j)\), the heat current S is computed as \cite{SPTC}:
 \begin{equation}
   S_{ij}^{\alpha}(m,n) = \sum_{\beta} \frac{e^{\alpha m}_{i} e^{\beta n}_{j} \phi^{\alpha \beta}_{ij}}{\sqrt{m_i m_j}} \Delta R_{ij}^{\alpha}  
 \end{equation}
 
\noindent where \(e^{\alpha m}_i\) is the component of the eigenvector for atom \(i\) in mode \(m\), and \(\phi^{\alpha\beta}_{ij}\) is the mass-weighted force constant tensor. The outputs include the squared magnitude \(S_{ij}^2\), and Cartesian components \(S_{ij}^{x}\), \(S_{ij}^{y}\), and \(S_{ij}^{z}\), stored in binary files for efficient access in the final SPTC calculation.

\subsection{Computing SPTC}
\label{subsec:SPTCcalculation}

The final module, \textsc{sptc\_calculation.cpp}, constructs the site-projected thermal conductivity tensor based on the precomputed heat current tensor and phonon mode data. Using the inputs from Sections~\ref{subsec:Harmonic} and~\ref{subsec:SPTCpreprocessing}, the code assembles the thermal transport matrix \(\Xi_{ij}\), as defined in Equation~\ref{eq:Gamma_def}, for a user-specified list of frequencies.

The core of the computation is the construction of the  transport ($\Xi$) matrix following Equation \ref{eq:Gamma_def}
with a defined width for the Lorentzian broadened delta function. The script then loops over a list of user-defined frequencies and accumulates thermal coupling strengths from all relevant mode pairs. Each entry in \(\Xi\) corresponds to energy transfer between two atomic sites mediated by vibrational coupling, and can be projected into Cartesian co-ordinates; $\Xi_{ij;\alpha}$.  Once \(\Xi\) is constructed, the thermal activity is projected onto atomic sites according to Equation \ref{eq:SPTC} and summed across all sites to yield the total conductivity, as in Equation \ref{eq:kappa_zeta}.

The \(\Xi\) matrix is optionally diagonalized to obtain its spectral structure and determine the modes of thermal conduction. The matrix is diagonalized using \texttt{dsyevd}  \cite{lapack99}.  The eigenvalues and eigenvectors are stored for further modal and localization analysis. Outputs include frequency-resolved site-level conductivity, the full transport matrix, and spectral decompositions.

\subsection{Interatomic Interactions}
\label{subsec:Models}

All simulations utilized highly accurate Gaussian Approximation Potentials (GAP) for silicon~\cite{GAP_Si}, silicon-oxide \cite{gap_Silica}, and carbon~\cite{C}. Structural relaxations were performed using \texttt{LAMMPS}, with an energy convergence criterion of $\Delta E = 10 \times 10^{-6}$ eV. The $DM$ was constructed via finite displacements of 0.05~\AA{} along each Cartesian direction. The SPTC method is easily adapted to \textit{ab initio} interactions. We have also demonstrated compatibility with \texttt{VASP}, although the computation of force constants in this case is significantly more demanding.

\section{\label{sec:Application_aSi} Amorphous Silicon}

In this section we apply the method to a-Si both to test and understand the method and also because it is a classic disordered system, explored in a vast experimental and theoretical literature \cite{Lewis,Street}.

We employ structurally credible a-Si models generated with the Wooten–Winer–Weaire (WWW)\cite{WWW} approach, as implemented by Djordjevi\ifmmode \acute{c}\else \'{c}\fi{}, Thorpe, and Wooten \cite{DTW}, and studied in Ref. \cite{IGRAM}. Systems containing 64, 216, 512, and 4096 atoms were analyzed to test our method against previous studies, track size effects and explore new features of TC in a-Si. The models were well relaxed using the same "annealing schedule" and all atoms were four-coordinated.  

\subsection{SPTC Representation of Locon, Propagon, and Diffuson Regions of the Vibrational Spectrum}\label{subsec:loconsPropagonsDiffusons}

AF classifies vibrational modes in disordered systems into three distinct categories: \emph{propagons}, \emph{diffusons}, and \emph{locons}~\cite{AF_LPD2,AF_LPD1}. \textit{Propagons} are low-frequency, wave-like modes analogous to acoustic phonons in crystals; they retain coherence over long distances and efficiently transport energy. However, in supercell calculations, even for large models, propagons are sparsely sampled, hence only propagons commensurate with the cell are computed.

\begin{figure}[!t]
    \centering
    \includegraphics[width=\linewidth]{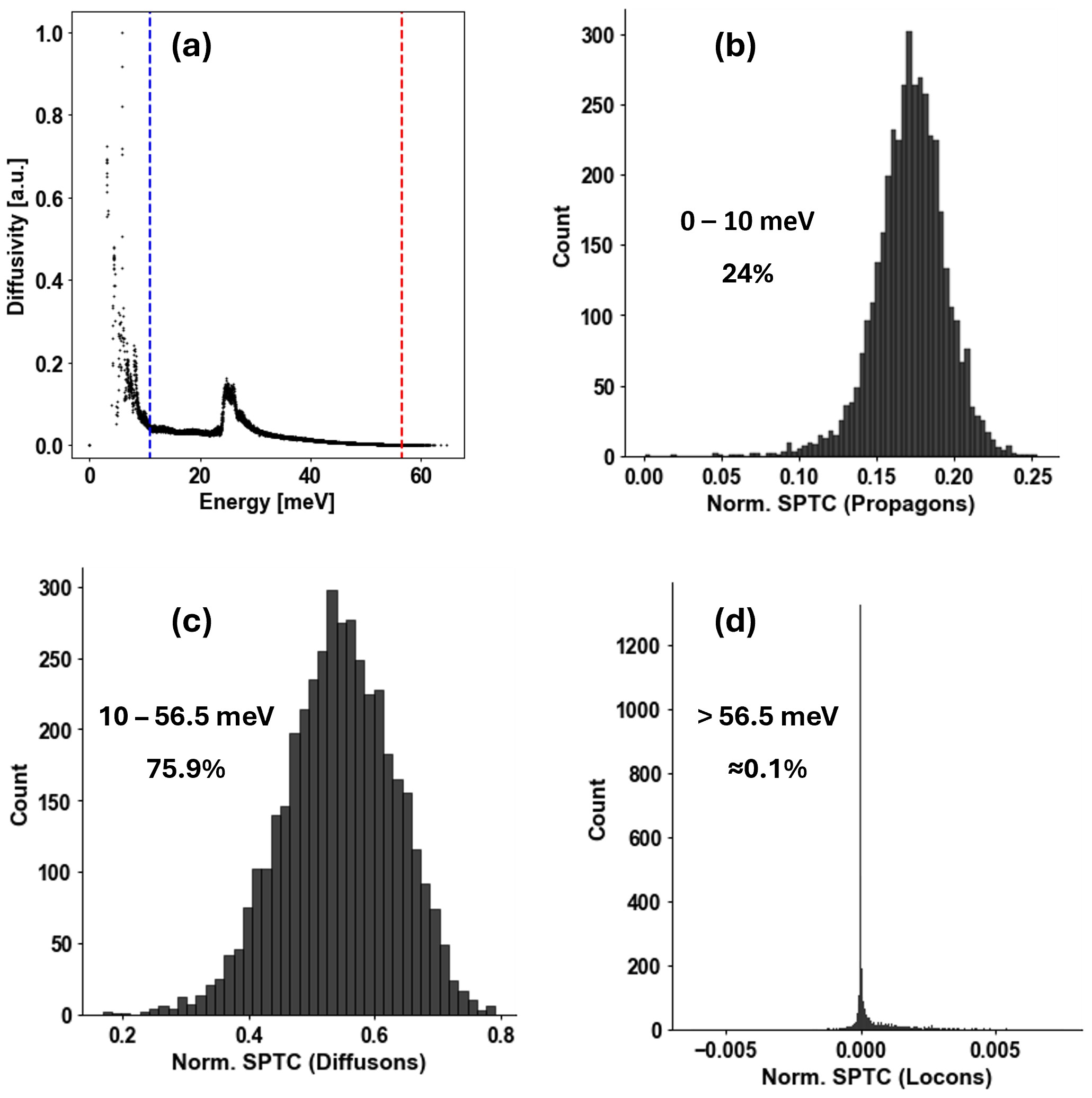}
    \caption{Vibrational mode analysis for the 4096-atom a-Si model. (a) Thermal diffusivity vs. vibrational energy, delineating energy regions for propagons (left of blue dashed line), diffusons (middle), and locons (right of red dashed line near the mobility edge). SPTC distributions from (b) propagons, (c) diffusons, and (d) locons. Insets indicate their energy ranges and percentage contributions to total SPTC. Similar plot for the 512-atom a-Si model is provided in Figure \textcolor{blue}{S2}.}
    \label{fig:aSi4096_modes}
\end{figure}

\textit{Diffusons} are spatially extended but non-propagating modes characterized by randomized eigenvectors without well-defined phase relationships. These modes dominate the vibrational spectrum of amorphous materials and contribute substantially to thermal conductivity via a non-ballistic, diffusive mechanism \cite{DiffusonCoherence}. \textit{Locons}, on the other hand, are highly localized vibrations, typically confined to small clusters of atoms and largely decoupled from the surrounding network. While their direct contribution to thermal transport is minimal, they can influence localized energy dissipation and anomalous thermal behavior. Following AF, we compute the thermal diffusivity \( D_n \) as~\cite{AF_LPD2}:

\begin{equation}\label{eq:diffusivity}
    D_n = \frac{\pi V^2}{3 \hbar \omega_n^2} \sum_{m \ne n} |S_{nm}|^2 \delta(\omega_n - \omega_m),
\end{equation}

\noindent where \( V \) is the system volume, \( \omega_n \) is the frequency of mode \( n \), and \( S_{nm} \) are matrix elements of the heat current operator. The thermal diffusivity for the 4096-atom a-Si model is plotted as a function of vibrational energy in Figure~\ref{fig:aSi4096_modes}a. Blue and red dashed lines delineate the spectral ranges of the three vibrational mode types: propagons (\(0-10 \, meV\)), diffusons (\(10-56.5~meV\)), and locons (\( > 56.5~meV\)). A sharp decline in diffusivity near \(56.5~meV\) (\(\approx 456~cm^{-1}\)) marks the \emph{mobility edge}—the boundary between extended (diffuson) and localized (locon) modes, per the Allen-Feldman (AF) framework. This classification is further supported by the vibrational density of states (VDOS) in Figure~\textcolor{blue}{S1} of the Supplementary Materials, which shows strong localization in the high-frequency locon regime.

The vibrational mode-projected SPTC was computed and its distribution is shown in histograms in Figures~\ref{fig:aSi4096_modes}b, \ref{fig:aSi4096_modes}c, and \ref{fig:aSi4096_modes}d for propagons, diffusons, and locons, respectively. Each histogram captures the spread of the SPTC values attributed to vibrational modes within a given energy range, with insets indicating the corresponding vibrational energy window and the maximum relative contribution (in percent) to the total system SPTC—defined as the calculated values without mode-projection. The propagon and diffuson distributions are approximately Gaussian, with peak SPTC contributions near \( 17 \,\%\) and \( 55 \,\%\), respectively. In contrast, the locon distribution is highly skewed, featuring a narrow peak at \( \approx 0.1 \,\%\) and a long tail, reflecting their highly localized nature. This analysis carried out for the 512-atom structure, shown in Figure \textcolor{blue}{S2}, exhibit similar distribution pattern.

\begin{figure}[!b]
    \centering
    \includegraphics[width=\linewidth]{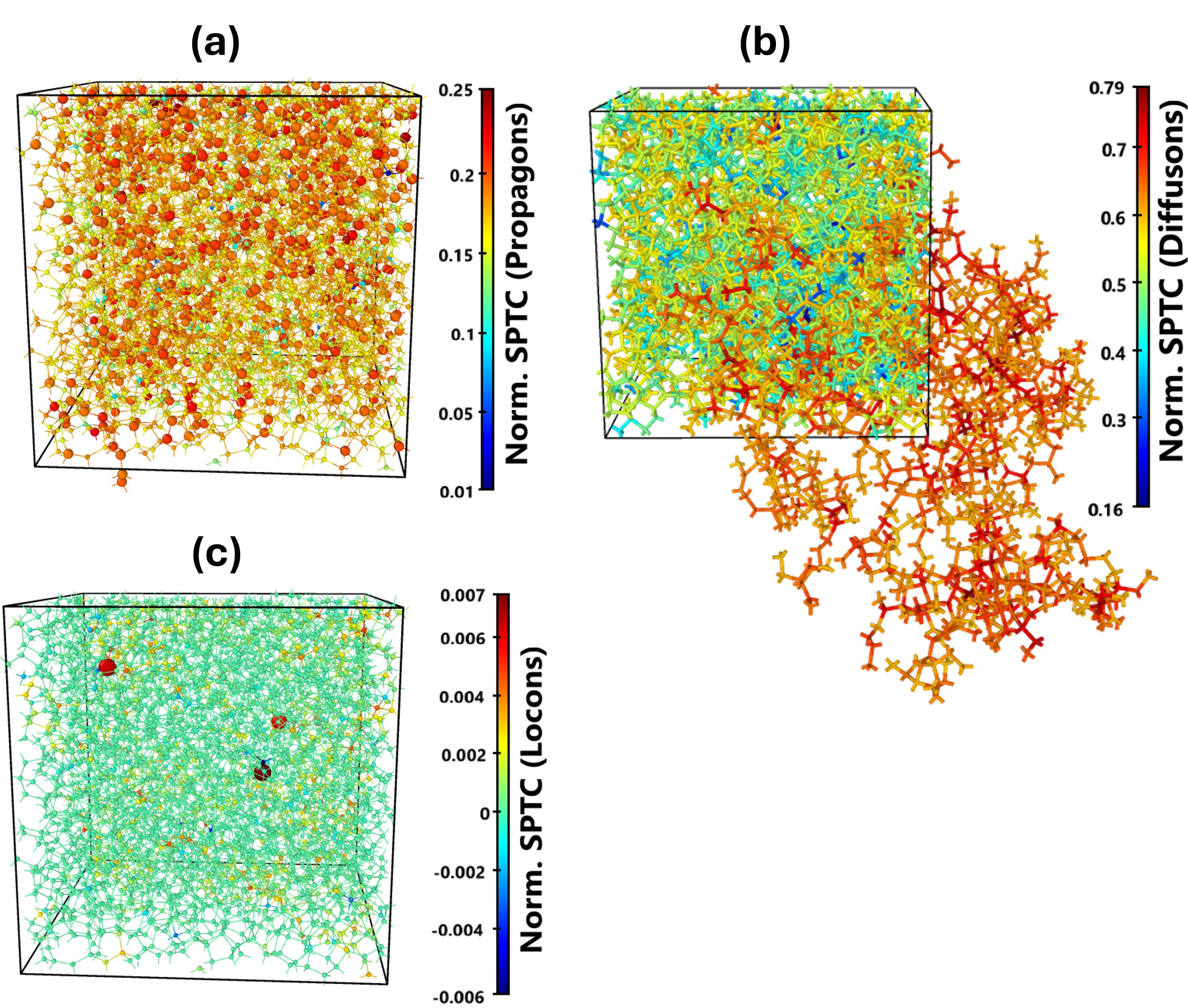}
    \caption{The contributions of (a) propagons, (b) diffusons, and (c) locons to the total SPTC are illustrated for the 4096-atom a-Si structure. The mode-projected SPTC is color-coded and its values are normalized by the total SPTC (i.e., without vibrational mode projection), so red regions correspond to the highest relative contributions within each mode class. Atoms contributing up to \( 75 \,\%\) of the mode's maximum SPTC (i.e., the high-SPTC range) are highlighted with increased radius, except in (b) where this is omitted for visual clarity. Atomic coordinates are unwrapped from periodic boundaries to better reveal the spatial connectivity of the top \( 75 \,\%\) contributors and their distribution throughout the simulation cell. Similar analysis for the  512-atom model is provided in Figure \textcolor{blue}{S3}.}
    \label{fig:fig_M_aSi4096_modeSPTC}
\end{figure}

Although propagon modes constitute only \(\approx 3 \,\%\) of the VDOS, they contribute disproportionately (\( 24 \,\%\)) to the total SPTC. These low-frequency, wave-like modes lie in the acoustic regime and are particularly sensitive to system size. While the SPTC formalism itself remains valid across system sizes, the projection method effectively captures only those propagons with wavelengths commensurate with the finite dimensions of the simulation cell. Consequently, the estimated propagon contribution may be systematically underestimated in limited-size models, a limitation further addressed in Section~\ref{subsec:FiniteSizeEffects}.

The spatial distribution of SPTC for propagon modes is relatively homogeneous across the structure. This is demonstrated in Figure~\ref{fig:fig_M_aSi4096_modeSPTC}a, where atoms contributing to the top \( 25 \,\%\) of the propagon SPTC are shown with increased radius. These high-contributing atoms are broadly distributed without significant clustering: 757 atoms fall within this high-SPTC range, yet the largest connected cluster contains only 18 bonded atoms. Additional details from the clustering analysis are provided in Table~\textcolor{blue}{S1}. The weak correlation with local structural motifs reinforces the interpretation that propagons facilitate a delocalized, system-wide mechanism of thermal transport.

Diffusons account for roughly \( 92 \,\%\) of the vibrational spectrum and contribute \( 75.9 \,\%\) to the total thermal conductivity. Spatial analysis of its SPTC distribution reveals significant clustering among high-contributing atoms (top \( 25 \,\%\)), as shown in Figure~\ref{fig:fig_M_aSi4096_modeSPTC}b. In this range, 1197 atoms are identified, of which 1093 form a single connected network through bonded interactions (see Table ~\textcolor{blue}{S1}). To illustrate this extended spatial connectivity, the atomic coordinates are unwrapped from periodic boundary conditions in the visualization. Diffuson modes with the highest SPTC values are found to correlate with local structural features, specifically bond lengths. The top 32 atoms ranked by SPTC exhibit an average bond length of 2.39~\AA, while the bottom 35 atoms (lowest SPTC) have a shorter average bond length of 2.30~\AA. For context, the overall mean bond length in the structure is 2.34~\AA. This bond-length dependence suggests that diffuson-mediated thermal transport is highly sensitive to local structural disorder and scattering strength. These observations support the interpretation of a filamentary character in heat conduction through amorphous silicon, a phenomenon that will be examined further in Section~\ref{subsec:SPTC_aSi}.

Finally, locons contribute a negligible \(\approx 0.1 \,\%\)to the total SPTC, with only three spatially isolated atoms accounting for up to \( 75 \,\%\) of the locon SPTC, as shown in Figure~\ref{fig:fig_M_aSi4096_modeSPTC}c. Notably, some of these contributions include negative SPTC values, which arise as numerical artifacts due to the ill-defined nature of the heat flux for highly localized modes. These modes exhibit minimal participation in heat transport, rendering their contribution to microscopic thermal conductivity essentially insignificant \cite{AF, Leitner}. 

The qualitative behavior of propagons, diffusons, and locons observed in the 4096-atom model is similar in the 512-atom a-Si structure. The spatial projection of the top \( 25 \,\%\) SPTC-contributing atoms for that case is provided in Figure~\textcolor{blue}{S3}, and the associated clustering data are summarized in Table~\textcolor{blue}{S1}.

\subsection{\label{subsec:FCM} Locality of the Force Constant Matrix $\implies$ Spatial Locality of SPTC}

The rapid spatial decay of the $FCM$ is a well-established feature that underpins many practical lattice dynamics calculations (\textit{e.g}., Ref.~\cite{Ordejon}). To gain analytical insight into this behavior, we consider in \ref{AppendixB} a generic pair potential of the Lennard-Jones type, with arbitrary powers $m$ and $n$ where $n < m$:

\begin{equation}
    V_{LJ}(r) = \left( \frac{A_m}{r} \right)^m - \left( \frac{B_n}{r} \right)^n,
\end{equation}

\noindent and derive the asymptotic form of the $FCM$ decay. As shown in \ref{AppendixC}, the SPTC summation converges for $n > 2$, confirming the necessity of "short-ranged" interactions for a well-defined thermal conductivity. 

\begin{figure}[t!]
    \centering
    \includegraphics[width=\linewidth]{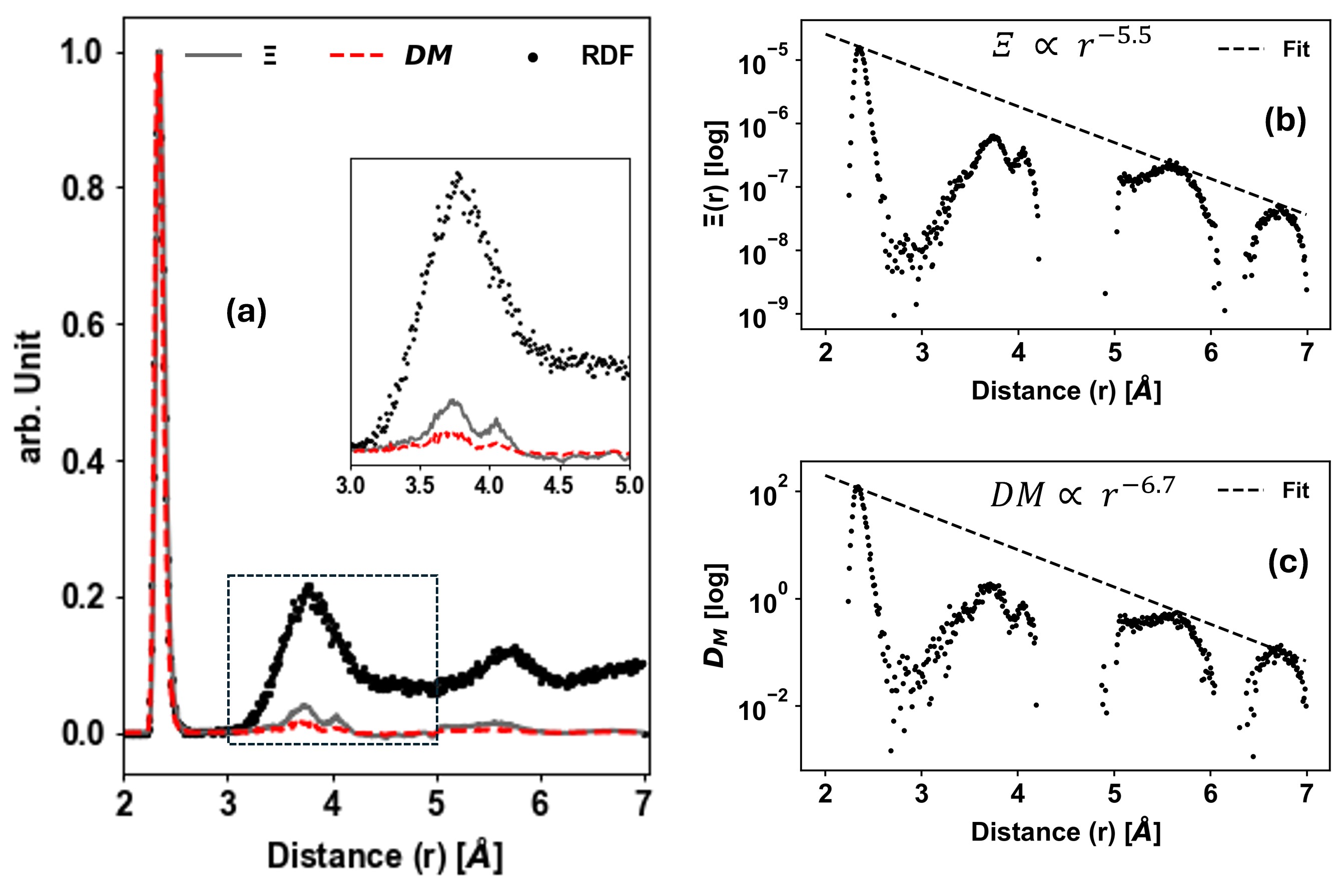}
\caption{Spatial convergence analysis of the 4096-atom amorphous silicon model. (a) Cell-averaged decay of $\Xi$ and $DM$ as a function of interatomic distance $r = |x - x'|$, shown alongside the radial distribution function (RDF, black) for reference. (b) Semi-logarithmic plot (log-scale on the y-axis) of $\Xi$ illustrating exponential decay behavior, with a linear fit (dashed black line) indicating a decay-rate power of 5.5. (c) Same as (b), but for $DM$, with a decay-rate power of 6.7.}
    \label{fig:fig_4096_Xi_DM_Decay}
\end{figure}

To assess this spatial decay empirically, we analyze the 512- and 4096-atom a-Si models. Figure~\ref{fig:fig_4096_Xi_DM_Decay}a shows the decay of  $\Xi$ and $DM$ as a function of radial distance \(r = |x - x'|\), averaged over all atomic sites, for the 4096-atom a-Si. For structural context, the radial distribution function (RDF) is also plotted, indicating the most probable neighbor distances. 

As expected, the RDF exhibits distinct peaks corresponding to the first (\(\approx 2.35\)~\AA), second (\(\approx 3.75\)~\AA), and third (\(\approx 5.8\)~\AA) nearest-neighbor shells. In contrast, neither $\Xi$ nor the $DM$ show significant features beyond the second/third neighbor shell. The inset in Figure~\ref{fig:fig_4096_Xi_DM_Decay}a highlights a weak second peak in $\Xi$, with an even fainter signal in the $DM$. To quantify the decay behavior, Figures~\ref{fig:fig_4096_Xi_DM_Decay}b and~\ref{fig:fig_4096_Xi_DM_Decay}c plot $\Xi(r)$ and $DM$ elements on a semi-logarithmic scale. Both quantities exhibit power-law decay, with exponents estimated to be 5.5 for $\Xi$ and 6.7 for the $DM$. The decay power for the $DM$ is consistent with quantum Monte Carlo simulations (Ref.~\cite{qMonteCarlo}). 

A corresponding analysis for the 512-atom a-Si model is provided in Figure~\textcolor{blue}{S4}, which shows consistent peak positions and decay profiles compared to the 4096-atom system. Notably, the decay power for $\Xi$ remains at 5.5, while the decay exponent for $DM$ decreased slightly to 6.1.

\begin{figure}[t!]
    \centering
    \includegraphics[width=\linewidth]{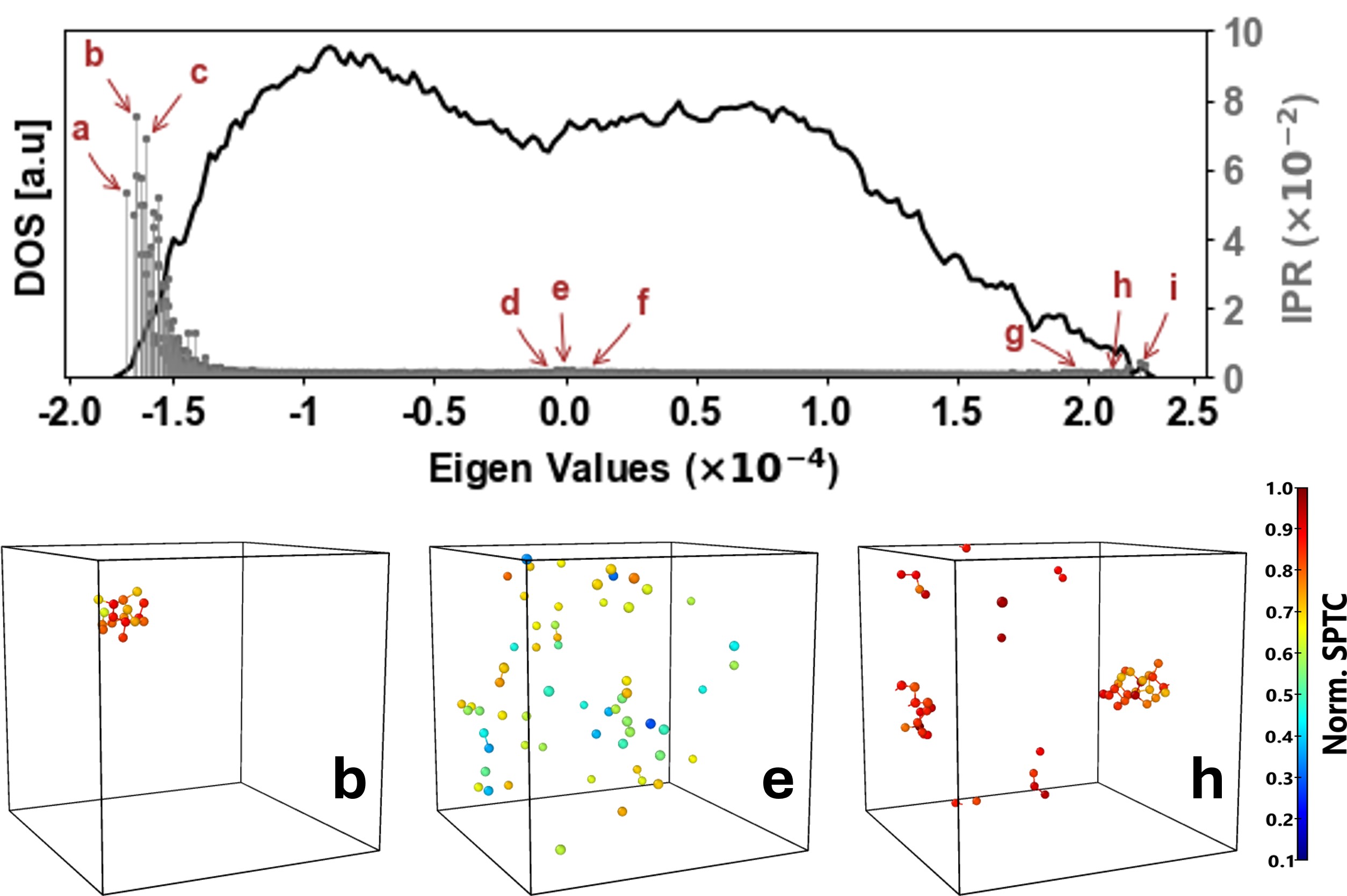}
    \caption{ The top panel shows the density of states (DOS; black) and inverse participation ratio (IPR; gray) of the $\Xi$ matrix for the 4096-atom a-Si structure. Nine representative eigenvalues (\textit{a}--\textit{i}) are selected from three distinct spectral regions and are highlighted. The atomic configurations corresponding to modes \textit{b}, \textit{e}, and \textit{h} are shown in the lower panel, with atoms color-coded by their SPTC contributions. Visualizations for the remaining highlighted eigenvalues, along with the DOS and IPR plots for the 512-atom structure, are provided in Figure~\textcolor{blue}{S5}.}
    \label{fig:fig_LocalizedEIGval}
\end{figure}

\subsection{The spectrum of $\Xi$}\label{subsec:spectrumOfXi}

The eigenvectors of $\Xi$ are helpful for identifying the most and least conducting parts of the network. We diagonalized $\Xi$ and present its density of states (DOS) and corresponding inverse participation ratio (IPR) in the top panel of Figure \ref{fig:fig_LocalizedEIGval}, for the 4096-atom a-Si model. We note that the eigenvectors ($\eta$) of $\Xi$ are mostly extended excepting those conjugate to the most negative eigenvalues, $\lambda$ (see Equation \ref{eqn:EigenValProblem}). To study this spatially, The atoms contributing to the (de)localization for nine representative eigenvalues  (labeled "\textit{a}--\textit{i}" in Figure \ref{fig:fig_LocalizedEIGval}) from three distinct regions were projected, and one mode from each region is shown in the lower panel of Figure \ref{fig:fig_LocalizedEIGval}, labeled \textit{b}, \textit{e}, and \textit{h}). The remaining six representative eigenvectors are provided in Figure~\textcolor{blue}{S5}.

The first region (\textit{a}, \textit{b}, \textit{c}), near the left band edge, exhibits spatially localized modes that form compact clusters (small blobs) composed of atoms with high SPTC (see colorbar for scale). In contrast, the second region (\textit{d}, \textit{e}, \textit{f}), where $\lambda \approx 0$ features extended modes, predominantly involving atoms with low to intermediate SPTC. The third region (\textit{g}, \textit{h}, \textit{i}), characterized near the right band edge, also includes interconnected atom clusters with high SPTC values, which we will correlate with our segmentation analysis in Section \ref{subsec:SPTC_aSi}. Moreover we find that a small fraction of the modes constitute most of the thermal conduction (these are modes near the high-$\lambda$ edge).

\begin{figure}[h!]
    \centering
    \includegraphics[width=\linewidth]{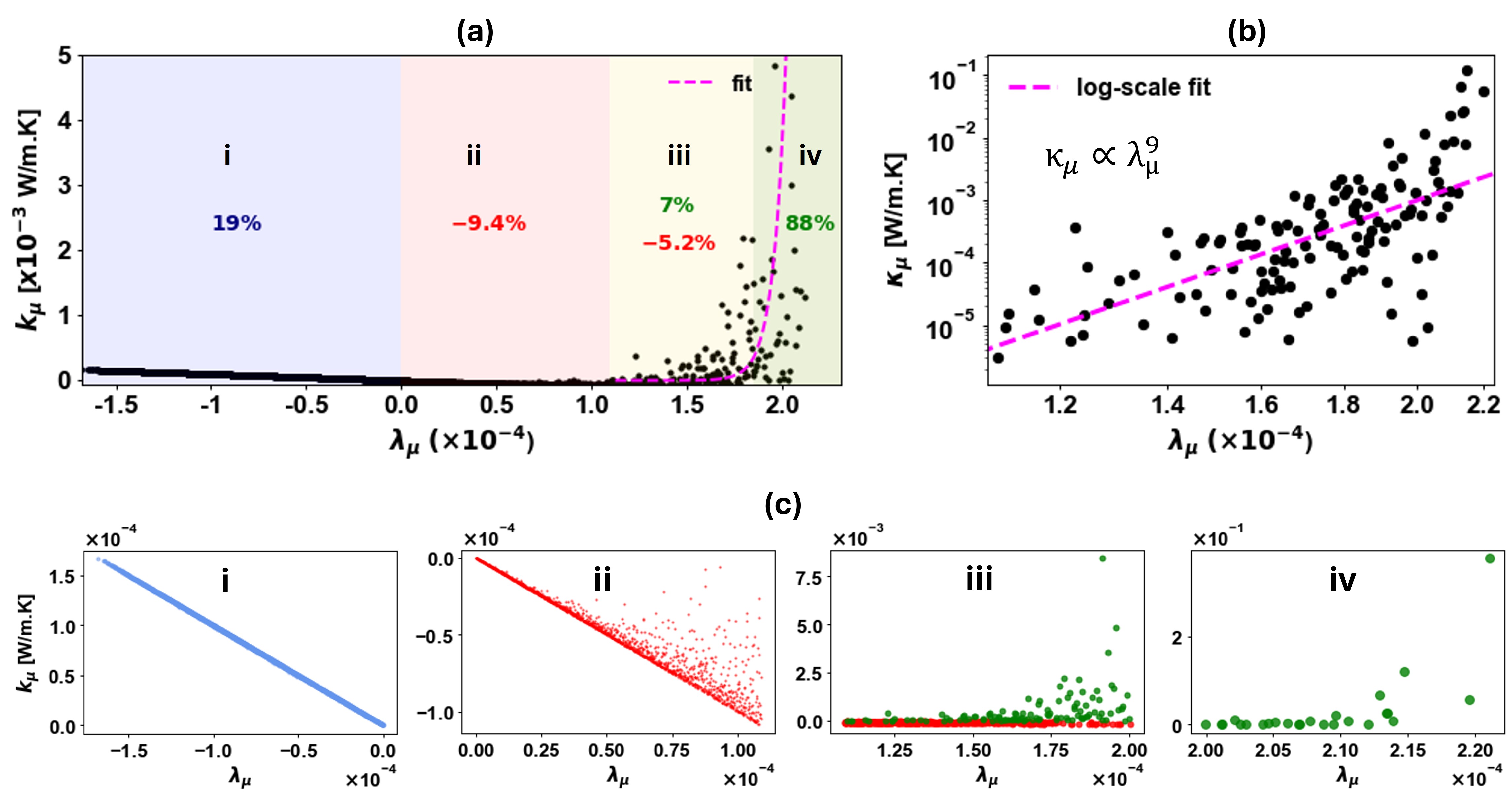}
    \caption{Mode-projected thermal conductivity, $\kappa_\mu$, for the 4096-atom a-Si structure. (a) The black scatter plot show $\kappa_\mu$ plotted against the eigenvalues ($\lambda_\mu$) of contributing eigenvectors, $\eta_\mu$. Colored bands highlight four distinct regions, i (blue) --- iv (green), of $\kappa_\mu$ contribution to the total conductivity $\kappa$ ($\approx$ 0.85~W/mK). The percentage contribution from each region is indicated by color, except in the yellow region (iii), which represents the combined contributions of the $-\kappa_\mu$ red (ii) and all $+\kappa_\mu$ green (iv) regions. (b) The power-law fit (magenta, dashed line) from the green and yellow region in (a) is replotted on a doubly logarithmic scale, revealing a decay power of 9.0. (c) Scatter plots i–iv show $\kappa_\mu$ versus $\lambda_\mu$ for the four regions, using their assigned colors for identification.}
    \label{fig:fig_kappaMu}
\end{figure}

To partly address this, we present the contributions of the eigenvectors $\eta_\mu$ (and their corresponding eigenvalues $\lambda_\mu$) to the mode-projected conductivity, $\kappa_\mu$ (see Equation \ref{eqn:kappa_mu}), as shown in Figure~\ref{fig:fig_kappaMu}a (black scatter). Four distinct regions (i–-iv) are identified based on their contributions to the total conductivity, $\kappa \approx 0.85$~W/mK:  (i; blue) The region with all negative eigenvalues ($\lambda_\mu < 0$), contributing \(19 \, \%\) to $\kappa$.  (ii; red) A set of low-positive eigenvalue modes (\(0 < \lambda_\mu \lessapprox 1.0 \times10^{-4}\)) which all yield negative values for $\kappa_\mu$, offsetting $\kappa$ by roughly \(9.4 \, \%\).  (iii; yellow) A transitional region (\(1.0 \times 10^{-4} < \lambda_\mu \lessapprox 1.9 \times 10^{-4}\)) containing both negatively (\(–5.2 \, \%\)) and positively (\(7 \, \%\)) $\kappa_\mu$ values. (iv; green) The dominant region (\( \lambda_\mu > 1.9 \times 10^{-4}\)), where all modes have positive values of $\kappa_\mu$ and contribute accounts for $\approx$ \(88 \, \%\) of $\kappa$.

A roughly power-law decay trend (magenta dashed line) is observed in the green-to-yellow region of Figure~\ref{fig:fig_kappaMu}a. When plotted on a doubly logarithmic scale (Fig.~\ref{fig:fig_kappaMu}b), this trend reveals a decay power of \( \approx 9\). Scatter plots in Figure~\ref{fig:fig_kappaMu}c (i–iv), color-coded by region, highlight the relationship between $\kappa_\mu$ and $\lambda_\mu$ for each contributing subset. Notably, the blue and red regions contain 2,100 and 1,438 modes, respectively. The yellow region comprises 422 negatively contributing modes and 111 positively contributing modes. In contrast, the dominant green region includes only 25 modes. These results suggest that thermal transport in a-Si is primarily governed by a small number of quasi-extended modes with large $\kappa_\mu$, while the majority of modes either contribute minimally or act destructively, offsetting the net conductivity.

This large $\kappa$ contribution from the modes in the large-$\lambda$ tail in the green region is reminiscent of the electronic version of this work in which we found that electronic conduction was determined by a tiny subspace of the full space of the eigenvectors of $\Gamma$ \cite{SPC1}. For the electronic case, $\Gamma$  is positive semi-definite, so that the non-contributing modes were concentrated at $\Lambda=0$, whereas for the thermal case only those at the positive spectral edge contribute.  For the electronic case, there was a “tail” in the density of states of $\Gamma$ near $\Lambda=0$ only for the case of a metal (Al), for the thermal case for a-Si, a narrow power-law tail is observed near the right band edge in the distribution of $\kappa_\mu$ (Figure \ref{fig:fig_kappaMu}a).

\subsection{Convergence of SPTC with Respect to Broadening and Spatial Cutoff}\label{subsec:SPTCconvergence}

In the SPTC formalism, the $\delta$  function in Equation~\ref{eq:Gamma_def} is approximated by a Lorentzian function \cite{AF_LPD2}:
\begin{equation}
    \delta(\omega) = \frac{\varepsilon}{\pi\left(\omega^2 + \varepsilon^2\right)}
\end{equation} 

\begin{figure}[t!]
    \centering
    \includegraphics[width=\linewidth]{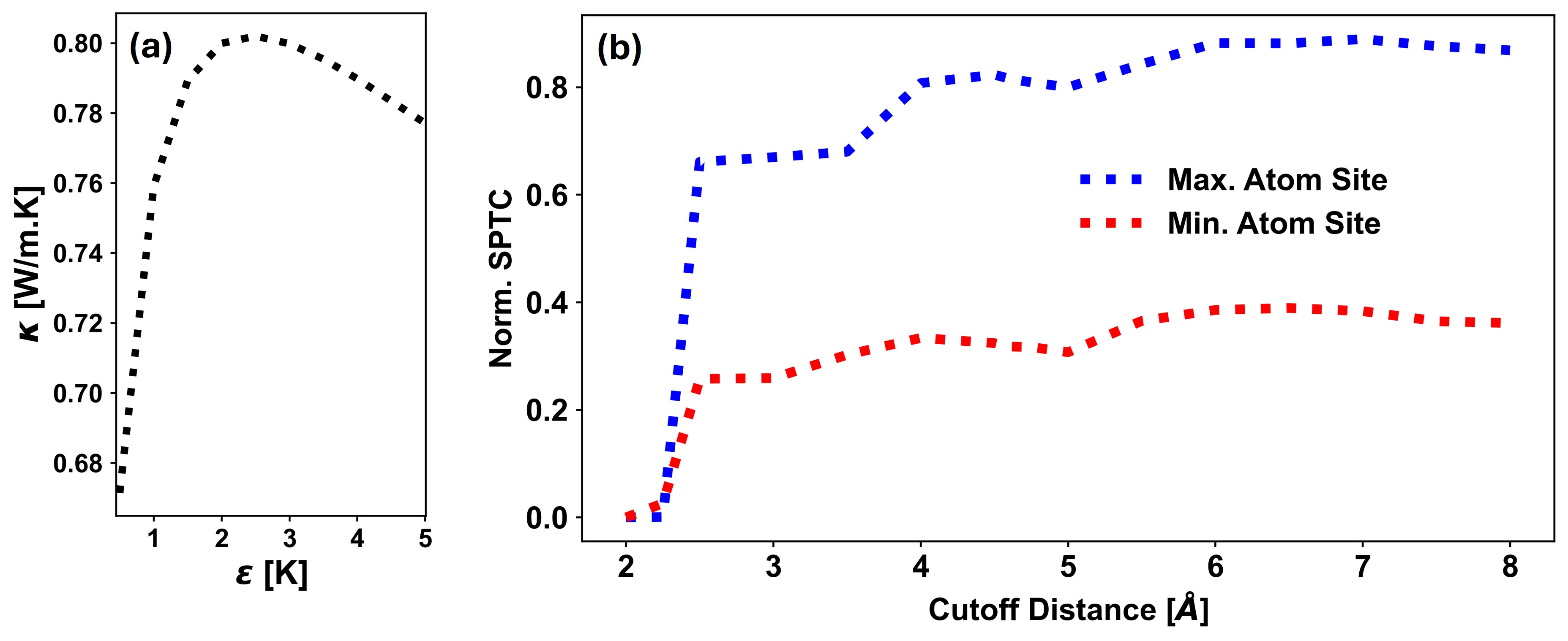}
    \caption{(a) Dependence of the computed thermal conductivity ($\kappa$) on the Lorentzian broadening factor ($\varepsilon$). (b) Convergence of site-projected thermal conductivity (SPTC) with respect to the spatial cutoff radius $R_c$, shown for the atomic sites with the highest (blue) and lowest (red) SPTC values. Results are shown for the 512-atom a-Si model.}
    \label{fig:fig_SPTC_Cutoffs}
\end{figure}

The influence of the Lorentzian broadening parameter, \(\varepsilon\), on the computed thermal conductivity ($\kappa$) is shown in Figure~\ref{fig:fig_SPTC_Cutoffs}a. We observe that $\kappa$ reaches a maximum near \(\varepsilon = 2\) K (in temperature units), and gradually decreases with further broadening. Based on this behavior, we adopt \(\varepsilon = 2\) K as the default value for all SPTC calculations in this work. This choice aligns with the broadening factor $\eta$ used in previous studies by Allen and Feldman~\cite{AF_LPD2}.

In addition, the decay of  $\Xi(r)$ with interatomic distance enables spatial truncation radius, $R_c$, to limit the computational. For a-Si, a conservative choice of \(R_c \approx 7.0\)~\AA{} is sufficient to capture the relevant physics. This is demonstrated in Figure~\ref{fig:fig_SPTC_Cutoffs}b, which shows the convergence of the SPTC for atomic sites with the maximum (blue) and minimum (red) contributions in the 512-atom a-Si model. Both curves plateau within the range of 6--7~\AA, implying the adequacy of this spatial cutoff. Consequently, for a fixed tolerance (degree of convergence of the SPTC for a given site), one only need contributions to a given site $R$ due to sites within a radius $R_c$ of $R$. Thus, it is easy to compute the SPTC in a linear scaling fashion \cite{Paper7, vanderbilt,Daw,Paper79}, once the $DM$ has been diagonalized. 

\begin{figure}[t!]
    \centering
     \includegraphics[width=.8\linewidth]{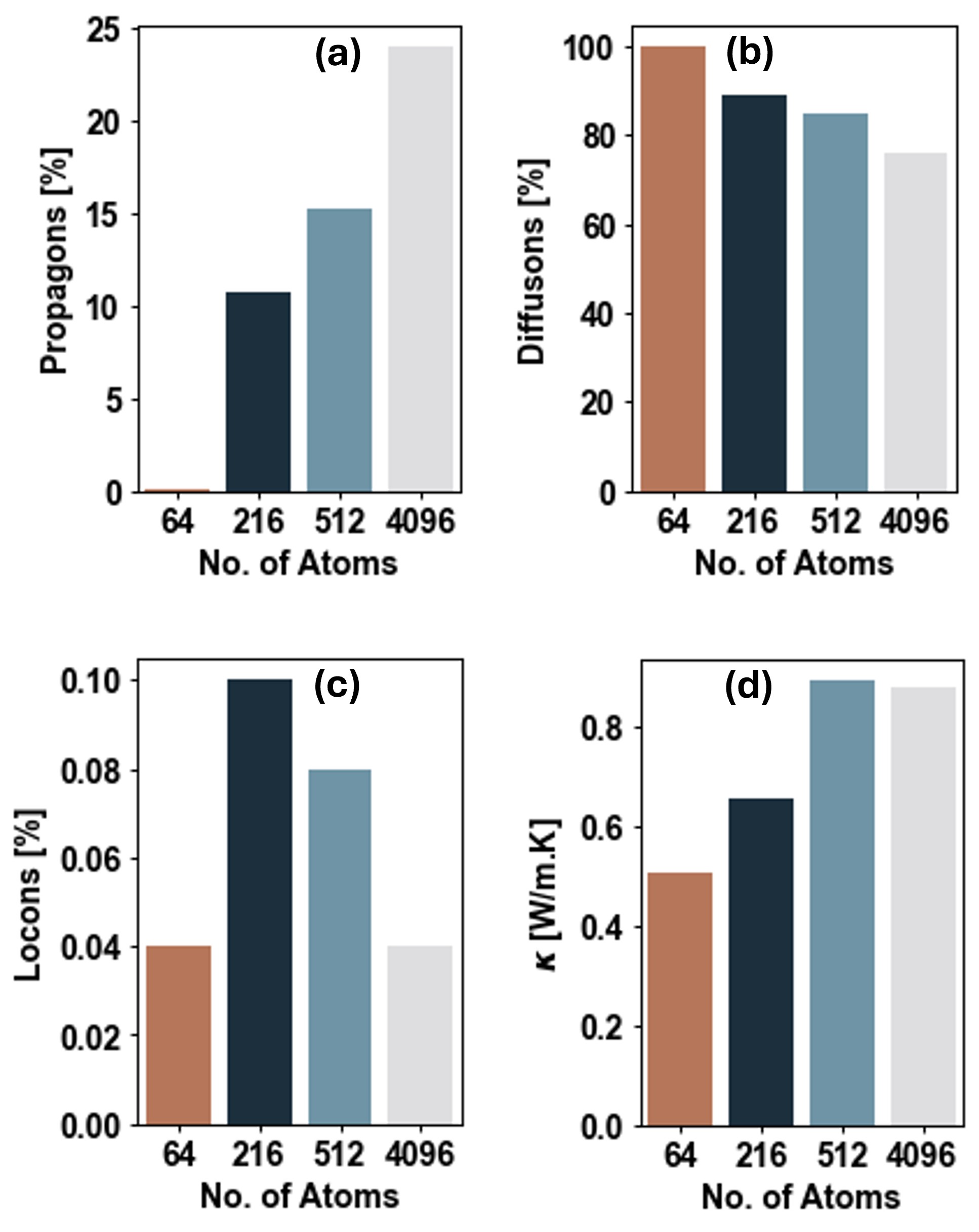}
    \caption{Size effects on SPTC calculations. Histograms showing the percentage contribution to the total SPTC from (a) propagons, (b) diffusons, and (c) locons for a-Si across four different system sizes. The corresponding trend in total thermal conductivity ($\kappa$) with increasing system size is shown in panel (d).}
    \label{fig:fig_aSi_sizeEffects}
\end{figure}

\subsection{Finite-size effects}\label{subsec:FiniteSizeEffects}

The effects of system size on SPTC of a-Si were assessed by comparing the fractions carried by propagons, diffusons and locons for supercells containing 64, 216, 512 and 4096 atoms.  In Table~\ref{tab:aSi_sizeEffects} and Figure~\ref{fig:fig_aSi_sizeEffects}a, the propagon contribution grows monotonically from \(0.15\,\%\) at \(N{=}64\) to \(24\,\%\) at \(N{=}4096\). This increasing trend of propagon contributing has also been reported for simulations carried out via the normal mode decomposition method \cite{Larkin2014}, and the contribution at \(N{=}4096\) is fairly close to the lower limit of \( 28 \,\%\) reported in frequency-domain thermoreflectance measurements for a-Si thin films \cite{regner}. 
 
\begin{table}[h!]
\centering
\scriptsize
\caption{Contributions of vibrational modes to SPTC and thermal conductivity ($\kappa$) in a-Si for different system sizes.}
\label{tab:aSi_sizeEffects}
\begin{tabularx}{\linewidth}{@{\extracolsep\fill}c c c c c@{}}
\toprule
\textbf{No. of Atoms} & \textbf{Propagons [\%]} & \textbf{Diffusons [\%]} & \textbf{Locons [\%]} & \boldmath{$\kappa$ [W/m$\cdot$K]} \\
\midrule
64    & 0.15  & 99.81 & 0.04 & 0.55 \\
216   & 10.73 & 89.17 & 0.10 & 0.65 \\
512   & 15.29 & 84.63 & 0.08 & 0.89 \\
4096  & 24.02 & 75.93 & 0.04 & 0.88 \\
\bottomrule
\end{tabularx}
\end{table}

Concomitantly, the contribution from diffusons decreases from \(99.8\,\%\) to \(75.9\,\%\) as system size increased (Figure~\ref{fig:fig_aSi_sizeEffects}b), mirroring the inverse propagon–diffuson correlation reported in time-domain thermoreflectance measurements on a-Si films~\cite{sizeEffects}. Locons remain below \(0.1\,\%\) for every size considered (Figure~\ref{fig:fig_aSi_sizeEffects}c), consistent with the AF picture that locons are spatially localised and therefore ineffective heat carriers~\cite{AF}. 

The overall thermal conductivity, \(\kappa\), rises from \(0.55\) to \(0.89\;\text{W\,m}^{-1}\text{K}^{-1}\) between 64 and 512 atoms, but then appears to saturate for 4096 atoms\footnote{It is of some interest that our results for diffusivity and thermal conductivity are in fairly close agreement with \textbf{FKAW}. Our models were of the same WWW-type, but we used the accurate GAP potential to handle the interatomic interactions. The results were similar enough to suggest that the details of the potential may not play a large role for thermal transport.} (Figure~\ref{fig:fig_aSi_sizeEffects}d).  A similar plateau emerges once the sample thickness exceeds the longest propagon mean-free paths in both  thermoreflectance experiments \cite{sizeEffects} and non-equilibrium molecular-dynamics simulations of a-Si~\cite{Larkin2014}. This then suggest that total heat transport and SPTC is governed by a balance between long-range propagons—whose contribution is limited by boundary scattering once their mean free paths exceed the box size—and short-range diffusons, whose diffusive character dominates the residual conductivity when propagons are fully scattered \cite{Baroni1}.

\begin{figure*}[!t]
    \centering
    \includegraphics[width=.8\linewidth]{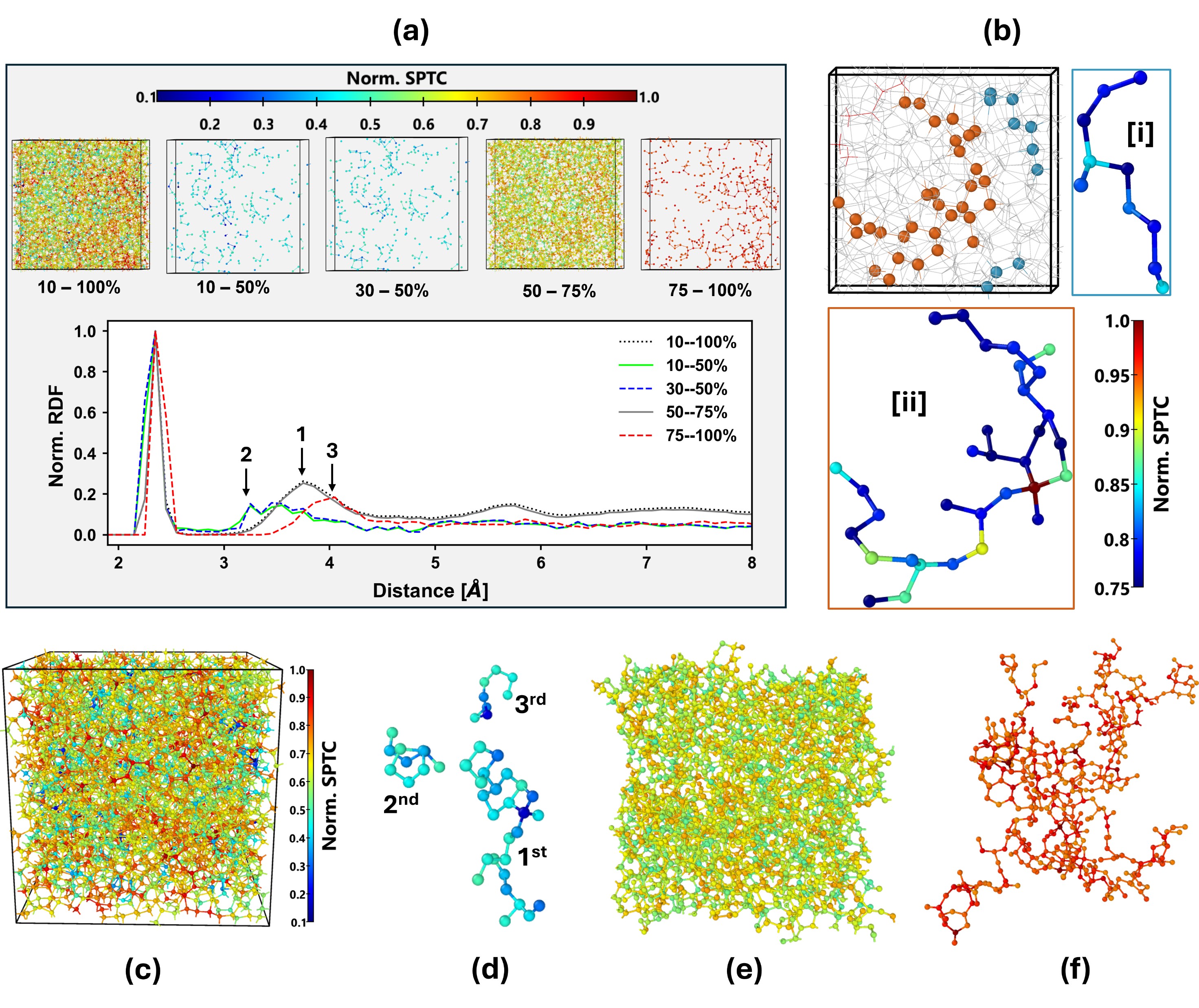}
    \caption{Site-Projected Thermal Conductivity (SPTC) analysis for two a-Si models containing 512 and 4096 atoms. (a) [Top] Spatial distribution of SPTC in the 4096-atom model segmented by range, with corresponding normalized radial distribution functions (RDFs). (b) High-SPTC filaments (SPTC  \( > 75 \,\%\)) in the 512-atom model, with blue and brown filaments highlighted in panels [i] and [ii], respectively. (c) Full SPTC map across atomic sites in the 4096-atom model. Visualization of (d) the first, second, and third low-SPTC filaments \( 10 - 50 \,\%\), (e) the largest mid-range SPTC cluster \( 50 - 75 \,\%\), and (f)  the longest high-SPTC filament in the structure. Clusters in (d–f) are color-mapped by SPTC value.}
    \label{fig:fig_SPTC_aSi}
\end{figure*}

\subsection{Anharmonic Effects}\label{subsec:Anharmonic}

To assess the anharmonic contributions to thermal conductivity, which are not captured by the harmonic approximation employed in the SPTC formalism (Equation~\ref{eq:kappa_zeta}), we computed the thermal conductivity of the 512-atom a-Si model using the Green-Kubo (GK) \cite{Green, Kubo} formalism from molecular dynamics (MD). In this approach, the thermal conductivity is obtained from the ensemble average of the heat current autocorrelation function (HCACF), given by~\cite{Green, Kubo, heatflux1, heatflux2}:

\begin{equation}
    \kappa_{\text{MD}} = \frac{1}{3Vk_BT^2} \int_0^{\tau} \left\langle \mathbf{J}(0) \cdot \mathbf{J}(t) \right\rangle \, dt,
    \label{eq:kappa_MD}
\end{equation}

\noindent where $V$ is the system volume, $T$ the temperature, and $k_B$ is Boltzmann's constant. The upper limit of integration, $\tau = 1$~ns, approximates the correlation time beyond which the heat flux autocorrelation function decays to zero. A Nosé–Hoover thermostat~\cite{nose, hoover} was used to thermalize and equilibrate the system at $T = 300$~K under microcanonical/constant volume (NVT) conditions, using a 1~fs time step. The equilibrated atomic velocities were subsequently used in a microcanonical (NVE) ensemble to obtain the HCACF.

The total thermal conductivity obtained from the Green-Kubo method was found to be 0.75~W/mK, while the harmonic phonon-based SPTC approach yielded a value of 0.88~W/mK. Both values are consistent with experimental measurements at room temperature~\cite{TC_aSi1, TC_aSi2}, indicating that anharmonic effects, though present, do not drastically alter the thermal conductivity in a-Si. This result supports the validity of the harmonic approximation as a reasonable framework for capturing the dominant mechanisms of thermal transport in a-Si.

\begin{table*}[!t]
\centering
\scriptsize
\caption{Cluster analysis for different SPTC ranges in the 512- and 4096-atom a-Si structures. Clusters are ranked from first (1$^\text{st}$) to eighth (8$^\text{th}$) largest in size. A cluster consists of at least two bonded atoms. $\overline{BL}$ denotes the average bondlength within each cluster.}
\label{tab:aSiFilaments}
\begin{tabularx}{\textwidth}{@{\extracolsep\fill}c c c *{8}{c} c@{}}
\toprule
\multicolumn{12}{c}{\textbf{512-atom Amorphous Silicon}} \\
\midrule
\textbf{SPTC Range [\%]} & \textbf{No. of Atoms} & \textbf{$\overline{BL}$ [\AA]} &
\textbf{1$^\text{st}$} & \textbf{2$^\text{nd}$} & \textbf{3$^\text{rd}$} & \textbf{4$^\text{th}$} &
\textbf{5$^\text{th}$} & \textbf{6$^\text{th}$} & \textbf{7$^\text{th}$} & \textbf{8$^\text{th}$} &
\textbf{Filament?} \\
\midrule
32--50       & 73   & 2.28 & 15 & 13 & 6 & 6 & 5 & 4 & 2 & 2 & Yes \\
50--75       & 376  & 2.34 & 375 & -- & -- & -- & -- & -- & -- & -- & No \\
75--100      & 63   & 2.39 & 32 & 10 & 6 & 3 & 2 & 2 & -- & -- & Yes \\
\addlinespace[1ex]
\multicolumn{12}{c}{\textbf{4096-atom Amorphous Silicon}} \\
\midrule
10--50       & 366  & 2.30 & 26 & 10 & 8 & 8 & 7 & 7 & 7 & 7 & Yes \\
50--75       & 2815 & 2.34 & 2792 & 3 & 2 & 2 & -- & -- & -- & -- & No \\
75--100      & 915  & 2.39 & 496 & 48 & 37 & 21 & 19 & 17 & 16 & 16 & Yes \\
\bottomrule
\end{tabularx}
\end{table*}

\subsection{SPTC for Amorphous Silicon}\label{subsec:SPTC_aSi}

In this section, we investigate the underlying structural configurations that may serve as preferential channels for thermal energy flow in a-Si. The observed transport behavior is predominantly governed by diffuson modes, as discussed in Section~\ref{subsec:loconsPropagonsDiffusons}.

Although the spatial distribution of SPTC in the a-Si models appears random at first glance, segmentation of the network into three conductivity-based sub-networks:low (\( 10 - 50 \,\%\)), intermediate (\( 50 - 75 \,\%\)), and high (\( > 75 \,\%\)), reveals emergent spatial organization. These sub-networks are visualized in the top row of Figure~\ref{fig:fig_SPTC_aSi}a for the 4096-atom configuration, with the full SPTC range spanning \( 10 - 100 \,\%\) (see enlarged view in Figure~\ref{fig:fig_SPTC_aSi}c). Notably, the lowest SPTC values remain finite: \( 32 \,\%\) and \( 10 \,\%\) for the 512- and 4096-atom system, respectively (see Table \ref{tab:aSiFilaments}). This indicates that atoms in low-conductivity regions still contribute non-negligibly to the overall thermal response. The selected SPTC thresholds are informed by characteristic features in the RDF (bottom row of Figure~\ref{fig:fig_SPTC_aSi}a), as well as by correlations between SPTC and local energetic and structural metrics (Figure~\textcolor{blue}{S6}a–b).

The RDF of the 4096-atom model reveals three distinct second-neighbor peaks (labeled "1", "2", and "3") associated with different SPTC intervals. Peak "1" corresponds to both the full system and the mid-SPTC range, centered near the expected second-neighbor distance of 3.75~\AA~\cite{IGRAM}. In contrast, peak positions shift to 3.45~\AA~and 4.05~\AA~for the low- and high-SPTC groups, respectively. These deviations provide physical justification for the segmentation thresholds. As shown in Table~\ref{tab:aSiFilaments}, a majority of atoms belong to the mid-SPTC category—approximately \(73\,\%\) and \( 69 \,\%\) for the 512- and 4096-atom models, respectively—consistent with the dominant contribution of this regime to the bulk structure and RDF profile.

The topology of these sub-networks is also of interest. Quantitative metrics for the sub-networks are provided in Table~\ref{tab:aSiFilaments}, including atom counts, cluster or filament sizes based on nearest-neighbor bond connectivity, and morphological classification. High-SPTC atoms tend to form extended, filamentary structures, as illustrated in Figure~\ref{fig:fig_SPTC_aSi}b, which highlights the two longest ``hot'' filaments in the 512-atom model, colored (i) blue and (ii) brown. In the 4096-atom model, the largest high-SPTC cluster (Figure~\ref{fig:fig_SPTC_aSi}f) contains 496 atoms and exhibits a more entangled, less linear morphology, reminiscent of polymer chains in high-density polyethylene~\cite{HDPE}. Additionally, the contributing atoms to the high-SPTC segment are primarily those associated with the extreme values of the eigenvalue spectrum, as discussed in Section \ref{subsec:spectrumOfXi}.

By contrast, low-SPTC atoms form small blobs, with the largest one containing only 26 atoms (see Figure ~\ref{fig:fig_SPTC_aSi}d). The mid-SPTC group represents the structural backbone: in the 4096-atom system, 2792 out of 2815 atoms form a single bond-connected network (Figure~\ref{fig:fig_SPTC_aSi}e). These observations suggest that high-SPTC filaments are embedded within the amorphous network and form a percolating substructure for thermal transport. These filaments are spatially extensive and effectively space-filling, extending throughout the simulation cell under periodic boundary conditions. This spatial continuity is evident in the visualization of the longest filament in the 4096-atom system, shown in Figure~\textcolor{blue}{S7}.

The average bondlength ($\overline{BL}$) for the sub-networks in Table \ref{tab:aSiFilaments} exhibit correlation with the SPTC ranges. $\overline{BL}$ increases progressively from the low- to high-SPTC sub-networks. In particular, the low-SPTC sub-networks exhibit shorter average bond lengths of 2.28 and 2.30~\AA\ for the 512-atom and 4096-atom models, respectively. By contrast, the mid- and high-SPTC sub-networks show a consistent $\overline{BL}$ of 2.34 and 2.39~\AA\ in both systems. The increasing bondlength in high-SPTC structures could indicate reduced local strain and enhanced vibrational coherence, consistent with their role in facilitating heat transport. 

There are two key observations regarding the bonding environment. First, the $\overline{BL}$ associated with the SPTC subnetworks closely match those obtained for diffuson modes, strongly suggesting that diffusons dominate the SPTC contributions. Second, the bond connectivity within the low- and high-SPTC subnetworks exhibits a structural self-correlation: short bonds preferentially connect to other short bonds, and similarly for long bonds, consistent with earlier findings~\cite{paper141, Inam}. Notably, prior works by Pan \textit{et al.} \cite{Urbach_Si} and Drabold \textit{et al.} \cite{paper164}, concerning Urbach tails \cite{Urbach}, showed that the valence tail states in a-Si are associated with connected short bonds, while the conduction tail states correspond to long bonds in a filamentary structure. Electronic conduction in the material is affected by the blob-filament characters of the Urbach tail because of its proximity to the Fermi level \cite{paper114, paper51}. Analogously, in the thermal case, we find that short bonds are predominantly involved in the low-SPTC subnetwork, whereas long bonds dominate the high-SPTC structures. This correspondence indicates a structural correlation between the local bonding environment and both the electronic and thermal transport properties of a-Si. Work is underway to determine to what degree \textit{the same} filaments or blobs contribute to electrical and thermal processes.

\begin{figure*}[!t]
    \centering
    \includegraphics[width=.8\linewidth]{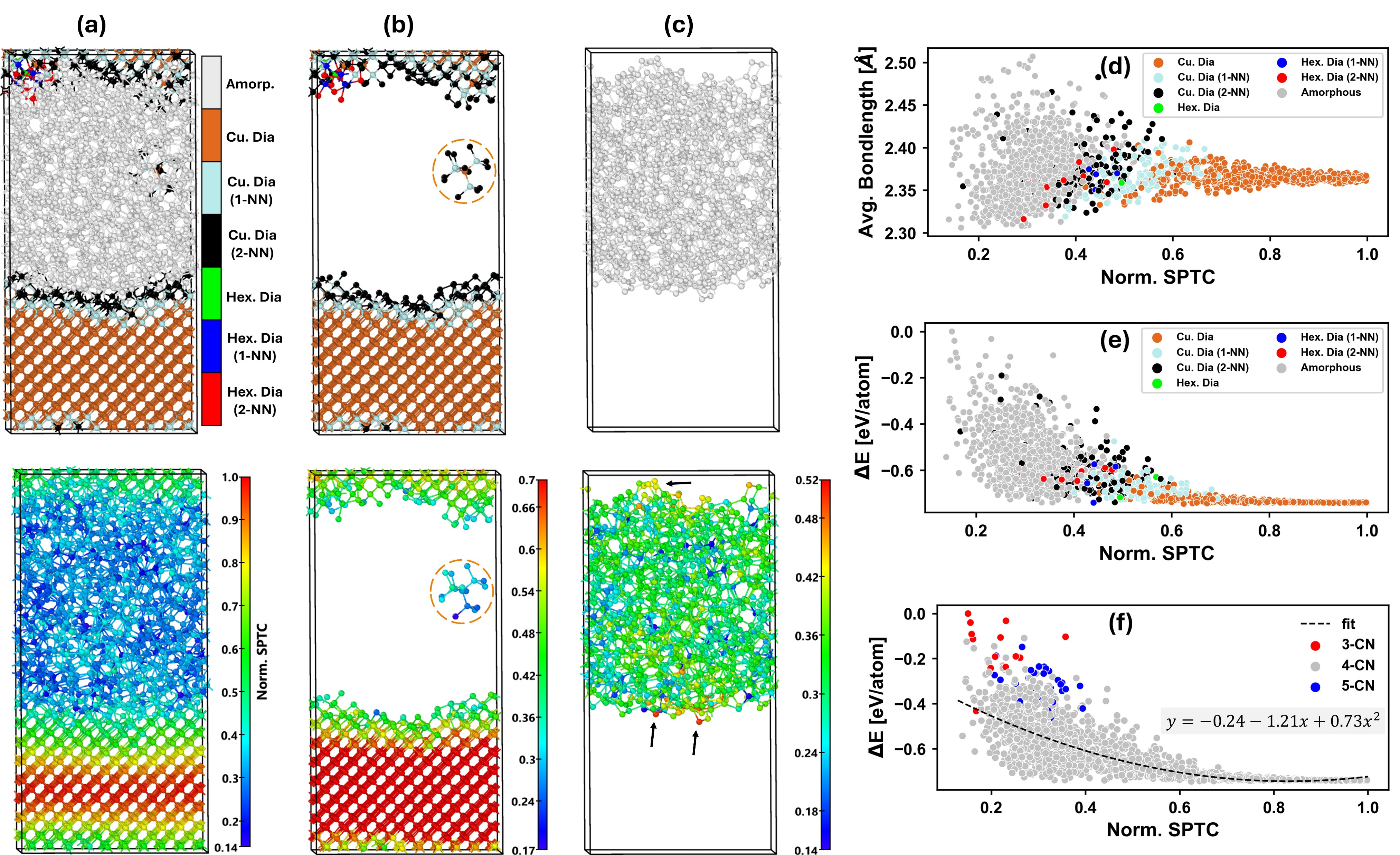}
    \caption{SPTC analysis of the amorphous/crystalline silicon sandwich structure. (a, top) Atomic configuration with color-coded structural classification: amorphous (Amorp.; gray), cubic diamond (Cu. Dia; brown), and hexagonal diamond (Hex. Dia; green), including first (1-NN) and second (2-NN) nearest neighbors (NN). The top panels of (b) and (c) delineate the crystal-like and amorphous regions, respectively. The lower panels of (a)–(c) show the normalized SPTC values per atom (blue: low, red: high).  (d) Relationship between SPTC, structure type, and per-atom average bondlength. The relationship between SPTC and phenomenological GAP site energy difference (\(\Delta E\)) is shown in (e) as a function of structure type and in (f) as a function of coordination number (CN). The black curve in (f) shows a quadratic fit, with the equation provided in the inset, and atoms are color-coded by coordination: 3-CN (red), 4-CN (gray), and 5-CN (blue).}
    \label{FIG:fig_SPTC_acSi}
\end{figure*}

\section{\label{sec:OtherApplication}Other Applications}

\subsection{A Crystal-Amorphous Interface in Silicon}\label{subsec:Bernstein}
Grain boundaries and interfaces play a critical role in determining the electronic and thermal properties of materials. Prior studies have shown that grain boundaries in silicon and tungsten can reduce SPTC by up to \( 30 \,\%\) \cite{SPTC, SPTCtungstein}. Here, we investigate the impact of amorphous–crystalline interfaces on thermal activity by analyzing a silicon sandwich structure composed of amorphous and crystalline domains, based on the model of Feldman and Bernstein \cite{Noam}. 

The initial bulk amorphous Si structure, consisting of 1000 atoms, was generated via molecular dynamics using the environment-dependent interatomic potential (EDIP). This was achieved by melting a crystalline system, quenching it to 1000 K, and annealing it over several million time steps. The composite amorphous/crystalline structure was constructed by duplicating the amorphous system along one periodic direction and extracting a 1222-atom slab, which was then combined with an 800-atom crystalline silicon slab oriented along the $\left<100\right>$ crystallographic direction. The resulting structure was relaxed and annealed at 1000 K under periodic boundary conditions. To improve energetic accuracy, the relaxed configuration was further annealed at 300 K under isobaric–isothermal conditions until the system reached a stable density of \(2.285 \, g/cm^3\), using the Gaussian Approximation Potential (GAP) for silicon \cite{GAP_Si} in \texttt{LAMMPS}\cite{lammps}. The transition from EDIP to GAP resulted in an energy reduction of \(\approx 0.036 \, eV\) per atom, with negligible changes in the atomic structure, suggesting that EDIP did a rather good job.

The top panel of Figure~\ref{FIG:fig_SPTC_acSi}a illustrates the amorphous/crystalline silicon sandwich structure, with atomic environments color-coded by structural type. Using the method from Ref. \cite{likeDiamond} as implemented in \texttt{OVITO}\textsuperscript{\textregistered} \cite{OVITO}, atoms were classified as amorphous Si (Amorp.; gray), cubic diamond (Cu. Dia; brown), or hexagonal diamond (Hex. Dia; green), along with their first (1-NN) and second (2-NN) nearest neighbors. Here, 1-NN (2-NN) designates atoms occupying lattice sites with at least one second (first) nearest neighbor not classified as crystalline. While the original crystalline region primarily exhibited a cubic diamond lattice, interface atoms partially transformed into the hexagonal diamond phase to accommodate the amorphous–crystalline transition. As summarized in Table \ref{tab:detailsNoam}, the number of atoms in the amorphous and cubic diamond regions decreased from 1222 and 800 to 1085 and 625, respectively. The SPTC profile in the sandwich structure demonstrates a decline in thermal activity from the cubic diamond region toward the amorphous phase.

\begin{table}[!b]
\centering
\scriptsize 
\caption{SPTC statistics across structural types in the amorphous–crystalline sandwich Si structure.}
\label{tab:detailsNoam}
\begin{tabularx}{\linewidth}{@{\extracolsep\fill}l c c c c@{}}
\toprule
\textbf{Structure} & \textbf{No. of Atoms} & \textbf{Max [\%]} & \textbf{Mean [\%]} & \textbf{Min [\%]} \\
\midrule
Cubic Dia.       & 625   & 100 & 81 & 42 \\
\quad 1-NN          & 151   & 70  & 53 & 34 \\
\quad 2-NN          & 146   & 61  & 42 & 25 \\
Hexagonal Dia.   & 2     & 57  & 53 & 49 \\
\quad 1-NN          & 4     & 49  & 45 & 43 \\
\quad 2-NN          & 9     & 48  & 38 & 29 \\
Amorphous           & 1085  & 52  & 31 & 14 \\
\bottomrule
\end{tabularx}
\end{table}

The normalized SPTC per atomic site is presented as a colormap in the lower panel of Figure~\ref{FIG:fig_SPTC_acSi}a, ranging from blue (low SPTC) to red (high SPTC). Atoms located in the bulk region of the cubic diamond phase exhibit the highest SPTC values ($\rightarrow$ 1), with a gradual decline observed toward the interface (red $\rightarrow$ yellow $\rightarrow$ green), where atoms display significantly lower SPTC values. To better distinguish the spatial distribution of SPTC in each phase, the crystalline and amorphous regions are indicated separately in Figures~\ref{FIG:fig_SPTC_acSi}b and \ref{FIG:fig_SPTC_acSi}c, respectively. In Figure~\ref{FIG:fig_SPTC_acSi}b, the colormap is capped at the maximum SPTC of cubic diamond 1-NN atoms (\(\approx 70\,\%\) of the bulk cubic diamond value; see Table~\ref{tab:detailsNoam}), yielding a spatial pattern that closely matches the structural classification. Similarly, in Figure~\ref{FIG:fig_SPTC_acSi}c, the maximum SPTC in the amorphous region reaches roughly \( 52 \,\%\) of the bulk cubic diamond value, with the highest values located near the interface, as indicated by black arrows. 

A strong correlation is observed between SPTC and the local bonding environment, as illustrated in Figure~\ref{FIG:fig_SPTC_acSi}d. Specifically, atoms with the highest SPTC are associated with interatomic distances converging around \(\approx 2.36\,\text{\AA}\)—the equilibrium bond length in crystalline silicon. Since structural identity is partially determined by interatomic distances (bond-angles also contribute), the high-SPTC regime is predominantly occupied by atoms in the cubic diamond phase, followed by contributions from hexagonal diamond configurations and their nearest neighbors. The amorphous region, by contrast, exhibits a significantly broader distribution with lower SPTC values, reflecting its disordered bonding topology.

This naturally raises the question: \textit{What physical quantity determines variations in SPTC?} The transition from crystalline to amorphous phases is accompanied by a gradual reduction in SPTC; however, this change is continuous—i.e., SPTC does not exhibit a discrete drop from 1 to 0 at the interface. It is therefore reasonable to hypothesize that the extent of local disorder in a material directly governs the disruption of heat transport. Since that structural features such as bond lengths vary smoothly in disordered (or at least non-crystalline) regions, establishing a direct correlation with SPTC is nontrivial. However, this challenge may be addressed by introducing local atomic energy as a surrogate descriptor, providing a physical link between structural disorder and observed variations in thermal conductivity.

\begin{figure*}[t!]
    \centering
    \includegraphics[width=.8\linewidth]{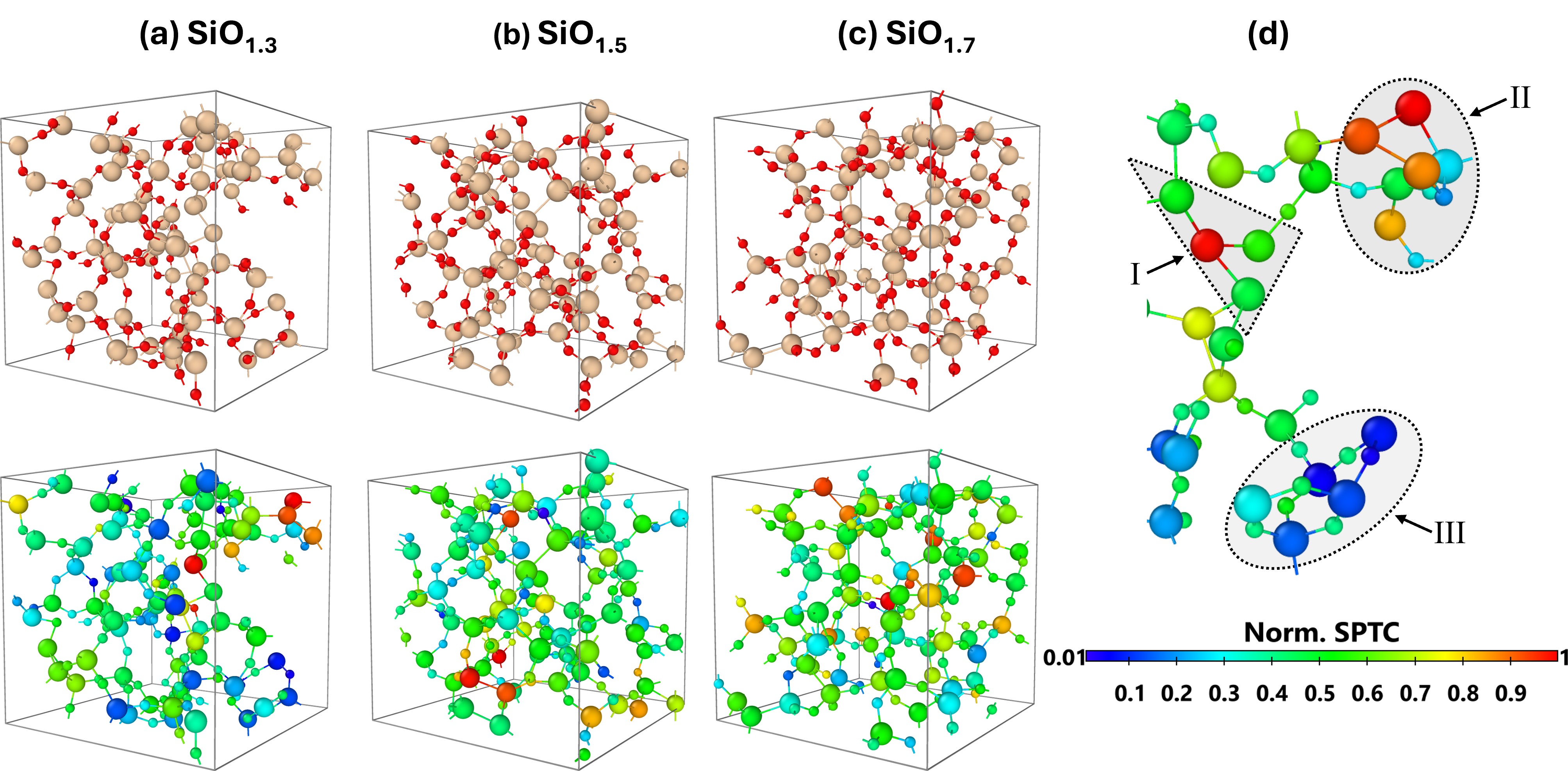}
    \caption{SPTC of silicon suboxides (SiO$_x$) with O concentrations of (a) 1.3, (b) 1.5, and (c) 1.7. Si and O atoms are shown as large brown and small red spheres in the top row. The bottom row depicts the normalized SPTC values using a colormap from blue (low) to red (high).  Thermal activities in Si-rich regions are highlighted in (d), with labeled regions I, II, and III in (e) indicating areas of interest.}
    \label{fig:fig_SPTC_SiO}
\end{figure*}

Figures~\ref{FIG:fig_SPTC_acSi}e and \ref{FIG:fig_SPTC_acSi}f illustrate the relationship between the difference in local energy  (\(\Delta E\), defined as the deviation of a particular site from the maximum energy, a fringe benefit of using the GAP potential) and SPTC. In Figure~\ref{FIG:fig_SPTC_acSi}e, atoms are categorized according to their structural classification, whereas in Figure~\ref{FIG:fig_SPTC_acSi}f, coordination number (CN) is used as the distinguishing metric. The data in Figure~\ref{FIG:fig_SPTC_acSi}e exhibit a trend: atoms within the cubic diamond lattice converge toward lower energy states and display higher SPTC values, while atoms in the amorphous phase occupy higher energy states with correspondingly lower SPTC. Atoms associated with other crystalline structure types fall within an intermediate energy/SPTC range. In contrast, when atoms are grouped solely by coordination number, as in Figure~\ref{FIG:fig_SPTC_acSi}f, the energy–SPTC relationship becomes less informative, particularly for 4-fold coordinated atoms. Although 3-fold and 5-fold coordinated atoms consistently exhibit higher energy and lower SPTC, the wide spread in energy/SPTC values within the 4-fold group obscures any definitive structural interpretation. This likely arises from the coexistence of both ordered and disordered atomic environments within this coordination class. 

However, a combined analysis incorporating both structural classification and coordination number provides a more nuanced understanding of how local structural disorder affects thermal transport, mediated through energy metrics. For example, the scattered black points in Figure~\ref{FIG:fig_SPTC_acSi}e that appear in the low-energy high-SPTC region correspond to "2-NN cubic diamond" labeled atoms but do not cluster with the main group of such atoms, which lie in an intermediate SPTC range. These outliers are highlighted by the brown dashed circle in Figure~\ref{FIG:fig_SPTC_acSi}b. Their anomalous energy and SPTC values arise from non-local effects from neighboring atoms surrounding this isolated cubic diamond substructure embedded within a predominantly amorphous region. Although their thermal and energetic signatures reflect this configurational isolation, such effects are not captured solely by structural classification or coordination number.

The black dotted curve in Figure~\ref{FIG:fig_SPTC_acSi}f denotes an emperical quadratic fit to the energy–SPTC relationship across all structural environments. However, when the analysis is confined to the amorphous regime (SPTC~\(\lessapprox 52 \,\%\)), a linear correlation emerges, characterized by a slope of \(-0.7\). This linear trend is consistent with observations from previously examined bulk a-Si models. Specifically, for the 4096-atom amorphous structure, the local atomic energy exhibits a linear dependence on SPTC with a slope of \(-0.41\), as shown in Figure~\textcolor{blue}{S6}a. An analogous linear relationship is also found between the average per-atom bond length and SPTC, with a slope of 0.10, as depicted in Figure~\textcolor{blue}{S6}b.

\subsection{A Silicon-Oxygen System}\label{subsec:Siox}
Next, we examine the thermal activity in a binary Si system, specifically non-stoichiometric silicon oxygen structures, i.e. silicon suboxide (SiO$_{x<2}$), a material identified as a promising candidate for secure storage devices due to its tunable electronic properties~\cite{SiOx}. The thermal active sites in three SiO$_{x<2}$ structures, obtained from Ref.~\cite{SiOx}, is shown in Figure~\ref{fig:fig_SPTC_SiO} for (a) SiO$_{1.3}$, (b) SiO$_{1.5}$, and (c) SiO$_{1.7}$, consisting of 70, 74, and 80 Si atoms, respectively. The top row illustrates atomic structures, with large brown spheres representing Si atoms and smaller red spheres representing O atoms, while the bottom row maps the SPTC per atom to a color scale ranging from blue (low) to red (high).

Unlike the electronic case discussed in Ref.~\cite{SiOx}, where active conduction pathways were identified in Si--Si connected regions, the thermal response in SiO$_{x<2}$ reveals spatially localized ``thermal hotspots.'' These are highlighted in Figure~\ref{fig:fig_SPTC_SiO}d, where Si-rich regions with prominent Si--Si bonding (labeled I and II) exhibit the highest SPTC values. In contrast, region III, similar to the silica-like environment of SiO$_2$, shows the lowest thermal activity, with Si atoms predominantly bonded to oxygen. Notably, the Si--Si bonded regions facilitate electronic and thermal transport in silicon suboxide.

\begin{figure*}[!t]
    \centering
    \includegraphics[width=.8\linewidth]{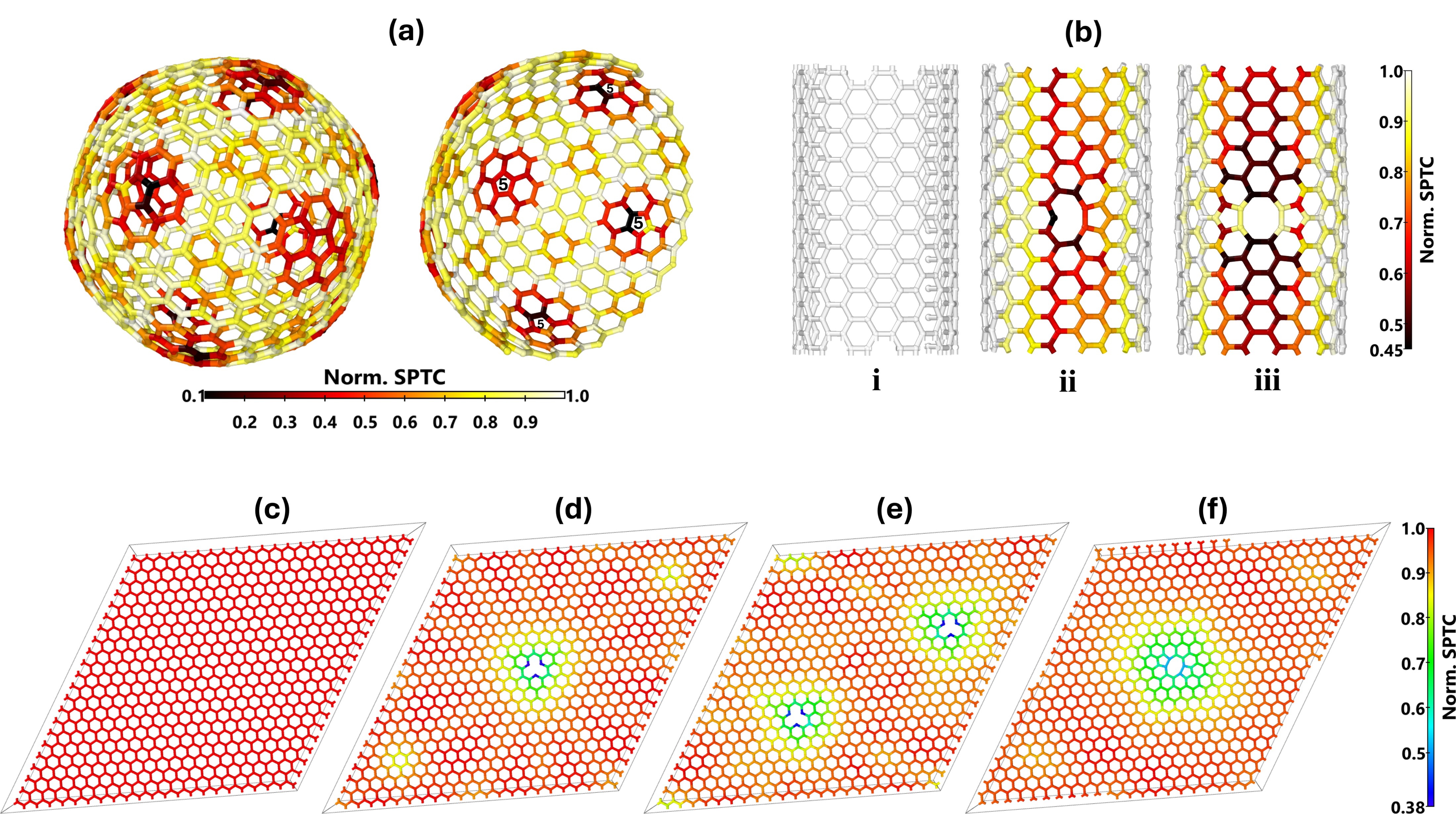}
    \caption{Thermal response visualizations for various carbon nanostructures: (a) A 720-atom fullerene model, highlighting 5-membered rings in one hemisphere using black labels. (b) A 400-atom carbon nanotube with (i) pristine structure, (ii) a single vacancy, and (iii) a vacancy pair. (c) A 968-atom graphene model, with vacancy configurations including (d) a single vacancy, (e) two spatially separated vacancies, and (f) a vacancy pair. Colorbars represent SPTC values for each structure.}
    \label{fig:fig_SPTC_Carbon}
\end{figure*}

\subsection{Carbon-based Materials}\label{subsec:carbon}

Shifting focus to carbon-based systems, we examine the thermal response of various carbon nanostructures and the influence of topological defects such as 5-, 7-, and 8-membered rings. The thermal activity in a 720-atom fullerene model is presented in Figure~\ref{fig:fig_SPTC_Carbon}a. A defining characteristic of fullerene is the incorporation of 5-membered rings (pentagons) between hexagons, introducing a positive Gaussian curvature that allows its stable and spherical geometry~\cite{BO}.One could interpret this as a closed 2D system. The SPTC projection onto atomic sites, as indicated by the colorbar, reveals that regions containing pentagons exhibit noticeably reduced thermal activity. 

However, one could argue that the reduced SPTC observed in fullerenes is a direct consequence of curvature rather than the presence of pentagons. To investigate this, we examined carbon nanotubes—another curved carbon allotrope. Unlike fullerenes, carbon nanotubes are composed exclusively of hexagonal rings, formed by rolling a graphene sheet into a cylindrical structure~\cite{aCNT}. As shown in Figure~\ref{fig:fig_SPTC_Carbon}b(i), the thermal activity across all atomic sites, including those on the curved surface, reaches the maximum SPTC value (normalized to 1). The introduction of topological defects, namely a single vacancy and a vacancy-pair, shown in Figures~\ref{fig:fig_SPTC_Carbon}b(ii) and (iii), respectively, leads to a reduction in thermal activity along the defect sites. In the single vacancy case, the SPTC in the defect region decreases to \(\approx 45 - 50 \,\%\) of the maximum bulk value. In contrast, the 5-8-5 defect configuration—arising from a vacancy pair—exhibits higher SPTC values (\(\approx 75 \,\%\)) on the atoms connecting the 5- and 8-membered rings.

It is well established that the positive Gaussian curvature introduced by a 5-membered carbon ring can be offset by the negative curvature associated with 7- or 8-membered rings~\cite{BO,aCNT, Mackay, TERRONES, Lenosky, aG2}. Consequently, topological connections between such ring types (e.g., 5--7 or 5--8 pairs) tend to favor the formation of more planar or low-curvature nano-structures. The 5--8--5 configuration shown in Figure~\ref{fig:fig_SPTC_Carbon}b(iii) exemplifies this curvature compensation mechanism, promoting localized planarity in the nanotube. We propose that the elevated SPTC values observed at the connecting atoms arise from this planar contribution. This hypothesis is further supported by analysis of a pristine 968-atom graphene model presented in Figure~\ref{fig:fig_SPTC_Carbon}c, as well as the effects of vacancy-induced defects shown in Figures~\ref{fig:fig_SPTC_Carbon}d--f. 

Analogous to the pristine nanotube, all atomic sites in the defect-free graphene lattice composed of hexagonal rings (Figure~\ref{fig:fig_SPTC_Carbon}c) exhibit maximal SPTC values, indicative of spatially uniform thermal activity. The introduction of a single vacancy (Figure~\ref{fig:fig_SPTC_Carbon}d) or spatially isolated vacancies (Figure~\ref{fig:fig_SPTC_Carbon}e) reveals a highly localized thermal response, consistent with the spatial decay of the $DM$ (see Section \ref{subsec:FCM}). This localization becomes particularly evident when comparing the SPTC associated with the 5-8-5 defect—a characteristic reconstruction of a divacancy—in both nanotube and graphene systems (Figure~\ref{fig:fig_SPTC_Carbon}f). In the nanotube, the inherent curvature allows the lattice to accommodate the defect while preserving structural coherence locally. In contrast, the flat geometry of graphene lacks such adaptive flexibility, leading to a substantial suppression of SPTC values—down to \(\approx 40 \,\%\) of the bulk maximum, at atomic sites connecting pentagonal and hexagonal rings. This reduction reflects the defect’s inability to contribute to in-plane structural stabilization, a role inherently fulfilled by the hexagonal framework in pristine graphene. The contrasting SPTC signatures between nanotubes and graphene thus highlight the strong dependence of thermal activity on local lattice topology, reinforcing the interpretation of SPTC as a fundamentally local probe of thermal behavior.

\section{\label{sec:conclucion} Conclusion}
We discuss the site-projected thermal conductivity (SPTC) formalism for identifying thermally active sites in materials and begins the process of identifying thermally active defect structures in materials. We also observe that a method providing local spatial information about thermal transport opens up the possibility of studying "thermal defects" in materials, those structural irregularities that limit or destroy heat conduction. It is not obvious that such thermal defects will be identical to electronic defects that produce localized states and/or affect electron transport. To the extent they turn out to be similar, it may be possible to properly understand the physical origin if the Wiedemann-Franz rule \cite{AshcroftMermin}, at least where the phonon part of the thermal conductivity is concerned. This work hints at that because of the importance of blobs and filaments to both Urbach tails and thermal transport. It may be that a unified understanding of both thermal and electronic defects is possible.

While the SPTC method offers valuable insights into microscopic heat transport, it is subject to certain limitations. First, the decomposition of the total thermal conductivity into atomic contributions—based on an analogy with Mulliken population analysis—is inherently non-unique, some reflecting ambiguity in assigning shared quantities to individual sites. Second, the present formulation relies on the harmonic approximation, thereby neglecting anharmonic effects that may influence thermal transport, particularly at elevated temperatures. Although these effects can, in principle, be incorporated in extended formulations, they are not captured in the current implementation. Finally, the lack of direct experimental probes for atomistic-scale thermal transport poses a challenge for validating SPTC predictions, limiting comparison to bulk thermal conductivity measurements and simulations.

\section*{Acknowledgments}
We thank Dr N. Bernstein for kindly providing crystal/amorphous Si interface model employed in this paper. We also thank Dr. R. M. Tutchton and Profs. S. Nakhmanson, J.-J. Dong, A. Demkov, and S. R. Elliott for helpful discussions.

\section*{Funding}
This work was supported by the US National Science Foundation under award MRI2320493 and the Office of Naval Research under grant N00014-23-1-2773. Additionally, C. U. acknowledges funding from the Laboratory Directed Research and Development program of Los Alamos National Laboratory under the Director's Postdoctoral Fellowship Program, project number 20240877PRD4. Los Alamos National Laboratory is operated by Triad National Security, LLC, for the National Nuclear Security Administration of U.S. Department of Energy (Contract No. 89233218CNA000001).

\appendix

\section{On decay of density matrices}\label{AppendixA}
To draw an analogy with electronic systems, Kohn attributed the spatial locality of quantum mechanics of electrons in the solid state to destructive wave-mechanical interference between single-particle electronic states that are nearly all delocalized, even for disordered systems \cite{Kohn1, Kohn2}. He named this "the principle of nearsightedness". Here, the spatial decay of $\Xi$, or equivalently $\rho(x - x')$ (see Equation \ref{eqn:Locality}), arises from the rapid decay of the $FCM$. In the electronic case, the density matrix $\rho_e$ decays exponentially in insulators, and in metals it becomes a power law decay\cite{Baym}: $\rho_e(x - x') \sim |x - x'|^{-2}, \quad \text{as } |x - x'| \rightarrow \infty$. Interestingly, the decay of $\Xi$ (or $\rho$) in amorphous silicon lies between these two limits: it is slower than the exponential decay characteristic of insulators but faster than the power-law decay observed in metals. This intermediate decay behavior accrues from the character of the vibrational modes in disordered systems.  

\section{The Spatial Decay of Force Constant Matrix}\label{AppendixB}
 To analytically examine the behavior of SPTC, we adopted a highly simplified model using an analytical interatomic potential. Specifically, we employed the Lennard-Jones (LJ) potential to describe the interaction between two particles. The LJ potential, commonly used to capture the essential features of intermolecular forces, is given by \cite{LJ}:
 \begin{equation}
 V_{LJ} \left( r \right) = \left( \left( \frac{A_m}{r} \right)^{m} - \left( \frac{B_n}{r} \right)^{n} \right)
 \end{equation}

 \noindent where $r$ is the distance between two atoms, and $n < m$. Considering the case where $n > 0$. The representation for LJ potential between two atoms, $i$ and $j$, with distance, $r_{ij} = r_{i} - r_{j}$, becomes:
 \begin{equation}
 \begin{aligned}
 U &= \frac{1}{2} \sum_{i \ne j} V_{LJ} \left( r_{ij} \right)
 \end{aligned}
 \end{equation}

Considering that:
 \begin{equation}
 \begin{aligned}
 \frac{\partial}{\partial r^{\alpha}_a} \left( \frac{A_m}{r_{ij}} \right)^{m} = - \frac{m}{A_m^2} \left( \frac{A_m}{{r_{ij}}} \right)^{m + 2} \left(r^{\alpha}_i - r^{\alpha}_j \right) \left( \delta_{ai} - \delta_{aj} \right)
 \end{aligned}
 \end{equation}

 \noindent the derivatives can be obtained as:
 \begin{equation}
 \begin{aligned}
 \frac{\partial U}{\partial r^{\alpha}_a} &= \sum_{i \ne a} \left( \frac{m}{A^2_m} \left( \frac{A_m}{r_{ia}} \right)^{m + 2} \left(r^{\alpha}_i - r^{\alpha}_a \right) - \frac{n}{B^2_n} \left( \frac{B_n}{r_{ia}} \right)^{n + 2} \left(r^{\alpha}_i - r^{\alpha}_a \right) \right)
 \end{aligned}
 \end{equation}

 For the case where $a \ne b$ (i.e., $r_{ab} \ne 0$), we have: 
 \begin{equation}
 \begin{aligned}
 \frac{\partial^2 U}{\partial r^{\alpha}_a \partial r^{\beta}_b} &= \frac{m}{A^4_m} \left( \frac{A_m}{r_{ba}} \right)^{m + 4} \left(r^2_{ba} \delta_{\alpha\beta} - \left( m + 2 \right) \left(r^{\beta}_b - r^{\beta}_a \right) \left(r^{\alpha}_b - r^{\alpha}_a \right) \right) \\
 &- \frac{n}{B^4_n} \left( \frac{B_n}{r_{ba}} \right)^{n + 4} \left(r^2_{ba} \delta_{\alpha\beta} - \left( n + 2 \right) \left(r^{\beta}_b - r^{\beta}_a \right) \left(r^{\alpha}_b - r^{\alpha}_a \right) \right)
 \end{aligned}
 \end{equation}

 Considering the relationship:
 \begin{equation}
     \left| \left(r^{\beta}_b - r^{\beta}_a \right) \left(r^{\alpha}_b - r^{\alpha}_a \right) \right| < r^2_{ba}  
 \end{equation}
 
 \noindent when $\alpha = \beta$, we have:
 \begin{subequations}
    \begin{align}
        \left| \frac{\partial^2 U}{\partial r^{\alpha}_a \partial r^{\beta}_b} \right| &< \frac{n \left( n + 1 \right)}{B^4_n} \left( \frac{B_n}{r_{ba}} \right)^{n + 4} r^2_{ba}\\
         &= \frac{n \left( n + 1 \right)}{B^6_n} \left( \frac{B_n}{r_{ba}} \right)^{n + 2}
    \end{align}
 \end{subequations}
 
\noindent and when $\alpha \ne \beta$, then the solution becomes:
  \begin{subequations}
    \begin{align}
     \left| \frac{\partial^2U}{\partial r^{\alpha}_a \partial r^{\beta}_b} \right| &< \frac{n \left( n + 2 \right)}{B^4_n} \left( \frac{B_n}{r_{ba}} \right)^{n + 4} r^2_{ba}\\
     &= \frac{n \left( n + 2 \right)}{B^6_n} \left( \frac{B_n}{r_{ba}} \right)^{n + 2} 
   \end{align}
 \end{subequations}
     
     Therefore, if we consider the case where $ n > 0 $, then:
    \begin{equation}
        \left| \frac{\partial^2U}{\partial r^{\alpha}_a \partial r^{\beta}_b} \right| < \frac{n \left( n + 2 \right)}{B^6_n} \left( \frac{B_n}{r_{ba}} \right)^{n + 2} 
    \end{equation}
    
 \section{The Convergence of spatial sums for SPTC}\label{AppendixC}
 Since both the $FCM$ and the $DM$ exhibit rapid spatial decay, the correlation function $\Xi(x - x')$ likewise decays for large values of $|x - x'|$. In Equation~\ref{eq:TC_Long}, the terms $e^{\alpha m}_{x}$, $e^{\beta n}_{x'}$, and $\frac{1}{\sqrt{m_{x} m_{x'}}}$ are independent of system size. Therefore, to understand the system-size dependence, we must focus on the denominator terms: 
 \begin{equation}\label{eq:flus_operator_Convergence_terms}
     \sum_{\gamma, x, x'} \phi_{x x'}^{\alpha \beta} \left( 0, \gamma \right) \left( \mathbf{r} \right)
\end{equation}

\noindent and their spatial behavior. First we simplify the Equation \ref{eq:flus_operator_Convergence_terms} using the following representations:
 \begin{subequations}
 \begin{align}
 \mathcal{G} &= \left|\phi_{x x'}^{\alpha \beta} \left( 0, \gamma \right)\right| \\
 \mathbf{r} &= \mathbf{R}_{xx'} + \mathbf{R}_{\gamma} \\
 \mathcal{G}(\mathbf{r}) &= \left|\phi_{x x'}^{\alpha \beta} \left( 0, \gamma \right) \left(\mathbf{R}_{xx'} + \mathbf{R}_{\gamma}  \right)\right|
 \end{align}
 \end{subequations}

 Next, we assume the distribution of the number of pairs $N$ from atom $x$ to atom $x^\prime$ in $\gamma^{th}$ cell are uniform for large $\mathbf{r}$, and we represent the the uniform distance as $\bar{r}$. The number density of pairs, $\rho$, relates to $N$ as:
 \begin{equation}
      dN = 4 \pi r^2 \rho dr
 \end{equation}
 
Referencing the finite terms for atom $x$ as $\mathcal{M}_x$, the size of the term in Equation \ref{eq:flus_operator_Convergence_terms} can be interpreted as:
 \begin{subequations}\label{eq:flus_operator_Convergence}
 \begin{align}
 \sum_{\gamma, x, x'} \mathcal{G}(r) &< \sum_{\gamma, x, x'} \mathcal{G} \left| \mathbf{r} \right| \\
 &\approx \sum_{x} \left [ \mathcal{M}_x + \int_{\gamma, x'}^{(r > \bar{r})} \mathcal{G} \left| \mathbf{r} \right| d N \right] \\
 &< \sum_{x} \left[  \mathcal{M}_x +  \int_{\bar{r}}^{\infty} \frac{n \left( n + 2 \right)}{B^6_n} \left( \frac{B_n}{r} \right)^{n + 2} r 4 \pi r^2 \rho dr \right] \\
 &= \sum_{x} \left[  \mathcal{M}_x + \int_{\bar{r}}^{\infty} 4 \pi \rho \frac{n \left( n + 2 \right)}{B^3_n} \left( \frac{B_n}{r} \right)^{n - 1} dr \right] \\
 &= \sum_{x} \left[ \mathcal{M}_x - \frac{n \left( n + 2 \right)}{\left( n - 2 \right)} 4 \pi \rho \frac{1}{B^2_n}  \left. \left( \frac{B_n}{r} \right)^{n - 2} \right |_{\bar{r}}^{\infty} \right] \\
  &= \sum_{x} \left[ \mathcal{M}_x + \frac{n \left( n + 2 \right)}{\left( n - 2 \right)} 4 \pi \rho \frac{1}{B^2_n} \left( \frac{B_n}{\bar{r}} \right)^{n - 2} \right] 
 \end{align}
 \end{subequations}

 Note that Equation~\ref{eq:flus_operator_Convergence}f is only valid for $n > 2$. In this case, the numerator scales linearly with the number of atoms in the system (i.e., the sum over $x$ in Equation~\ref{eq:flus_operator_Convergence}f); however, this scaling is effectively canceled by the division by the system volume $V$ in Equation~\ref{eq:TC_Long}. As a result, for $n > 2$, the expression in Equation~\ref{eq:flus_operator_Convergence_terms} converges. Consequently, the thermal conductivity $\kappa$ in Equation~\ref{eq:TC_Long}, the spatial correlation function $\Xi(x, x')$ in Equation~\ref{eq:Gamma_def}, and the site-projected thermal conductivity, $\zeta(x)$, in Equation~\ref{eq:SPTC} are all well-defined and possess finite, convergent values.


\bibliographystyle{elsarticle-num}
\bibliography{SPTCmanuscript}

\end{document}


\title{SUPPLEMENTARY MATERIAL \\ Spatially Local Estimates of the Thermal Conductivity of Materials}

 \author[LANL]{C. Ugwumadu}
 \ead{cugwumadu@lanl.gov}
 \affiliation[LANL]{organization = Quantum and Condensed Matter Physics (T-4) Group, addressline = { Los Alamos National Laboratory}, city = {Los Alamos}, postcode = {87545}, state ={NM}, country={USA}}

\author[NQPI]{A. Gautam}
\ead{ag007122@ohio.edu}
\affiliation[NQPI]{organization = Department of Physics \& Astronomy, Nanoscale & Quantum Physics Institute, addressline = {Ohio University}, city = {Athens}, postcode = {45701}, state ={OH}, country={USA}}

\author[INPP]{Y. G. Lee}
 \ead{yl518521@ohio.edu}
 \affiliation[INPP]{organization = Department of Physics \& Astronomy, Institute of Nuclear & Particle Physics, addressline = {Ohio University}, city = {Athens}, postcode = {45701}, state ={OH}, country={USA}}

 \author[NQPI]{D. A. Drabold}
\ead{drabold@ohio.edu}

\date{\today}

\begin{abstract}
In this paper we describe a spatial decomposition of the thermal conductivity, what we name "site-projected thermal conductivity", a gauge of the thermal conduction activity at each site. The method is based on the Green-Kubo formula and the harmonic approximation, and requires the force-constant and dynamical matrices and of course the structure of a model sitting at an energy minimum. Throughout the paper, we use high quality models previously tested and compared to many experiments. We discuss the method and underlying approximations for amorphous silicon, carry our detailed analysis for amorphous silicon, then examine an amorphous-crystal silicon interface, and representative carbon materials. We identify the sites and local structures that reduce heat transport, and quantify these (estimate the spatial range) over which these "thermal defects" are effective. Similarities emerge between these filamentary structures in the amorphous silicon network which impact heat transport, electronic structure (the Urbach edge) and electronic transport. 
\end{abstract}

\begin{keyword}
thermal conductivity; molecular simulation; heat transport; SPTC; dynamical matrix
\end{keyword}

\maketitle

\section{\label{sec:Figures}Supporting Tables}

\begin{table}[!h]
\centering
\scriptsize
\caption{Cluster analysis of atoms contributing up to 75\% of the total SPTC from different vibrational mode classes—propagons, diffusons, and locons—in the 512- and 4096-atom amorphous silicon (a-Si) structures. Clusters are ranked from the first (1$^\text{st}$) to the eighth (8$^\text{th}$) largest by size. A cluster contains at least two bonded atoms. Diffusons exhibit the longest connectivity, involving 60\% of the top 25\% contributing atoms in the 512-atom model and 92\% in the 4096-atom model. Propagons follow with 27\% connectivity in the 512-atom model, but only 2.3\% in the 4096-atom model. In contrast, locons are spatially isolated, with only three atoms contributing up to 75\% of the mode's maximum SPTC in both models.}
\label{Stab:aSiModeFilaments}
\begin{tabularx}{\linewidth}{@{\extracolsep\fill}l *{9}{c} @{}}
\toprule
\multicolumn{10}{c}{\textbf{512-atom Amorphous Silicon}} \\
\midrule

			\textbf{Modes}  & \textbf{No. of Atoms}  & \textbf{1$^\text{st}$} & \textbf{2$^\text{nd}$} & \textbf{3$^\text{rd}$} & \textbf{4$^\text{st}$} & \textbf{5$^\text{st}$} & \textbf{6$^\text{st}$} & \textbf{7$^\text{st}$} & \textbf{8$^\text{st}$}   \\ 
            \midrule
			Propagons             & 111    & 31  & 16  & 5 & 4 & 3 & 3 & 2 & 2           \\
			Diffusons             & 82   & 49  & 15  & 4 & 2 & 2 & -- & -- & --            \\  
			Locons                & 3    & --  & --  & --   & -- & -- & -- & -- & --           \\ 
            
           \addlinespace[1ex]
            
            \multicolumn{10}{c}{\makebox[0pt]{\textbf{4096-atom Amorphous Silicon}}} \\
            \midrule
			Propagons          & 757 & 18  & 15  & 14   & 14 & 12 & 11 & 10 & 7           \\
			  Diffusons          & 1187   & 1093 & 10 & 5   & 3 & 3 & 2 & 2 & 2            \\  
			Locons                & 3    & --  & --  & --   & -- & -- & -- & -- & --           \\
            \bottomrule
		\end{tabularx}
\end{table}
\newpage

\section{\label{sec:Figures}Supporting Figures}

\begin{figure}[h!]
    \centering
    \includegraphics[width=.8\linewidth]{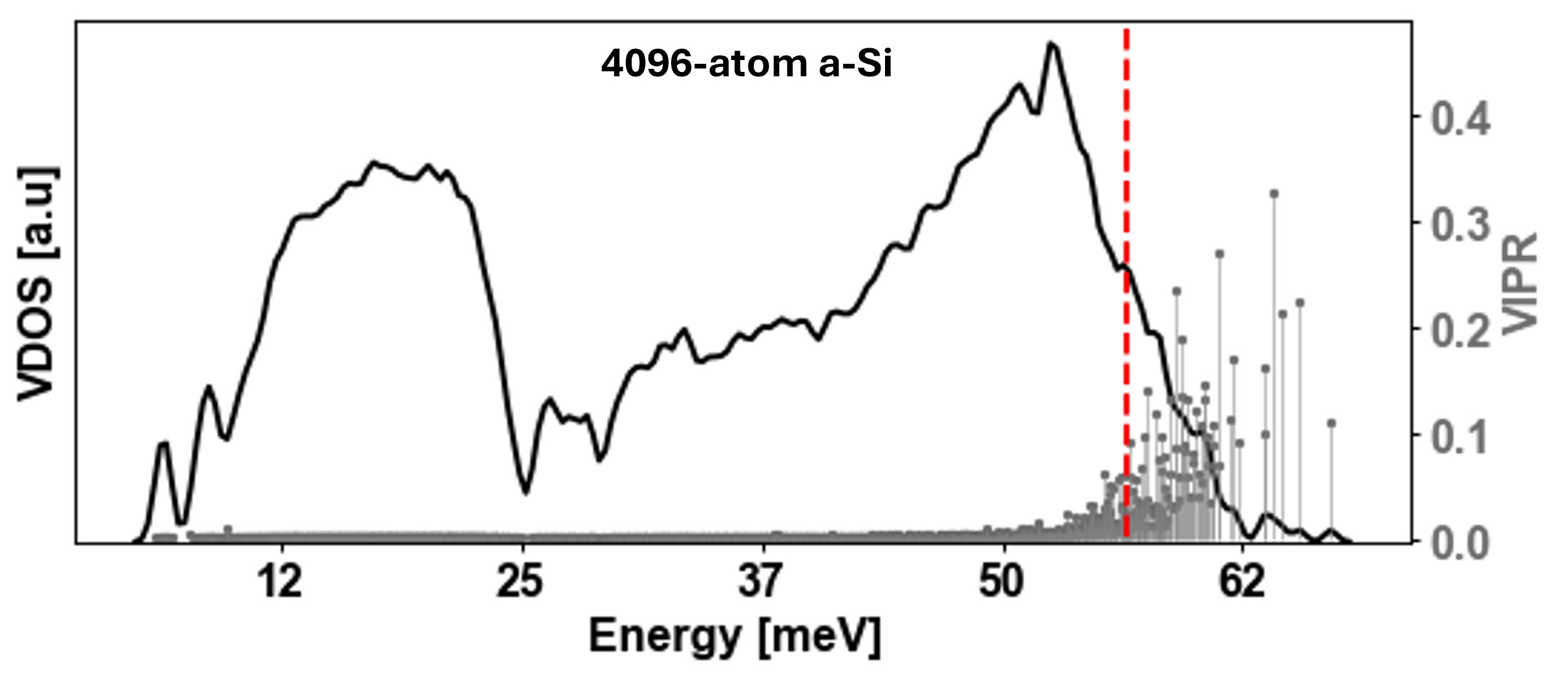}
    \caption{Vibrational density of states (VDOS, black) and vibrational inverse participation ratio (VIPR, gray) for the 4096-atom a-Si. The  red dashed line at  around 56.5 meV (450~cm$^{-1}$) denotes the approximate mobility edge, separating extended modes (propagons and diffusons) from localized modes (locons).}
    \label{fig:aSi_vdos_diffusivity}
\end{figure}

\begin{figure}[!h]
    \centering
    \includegraphics[width=\linewidth]{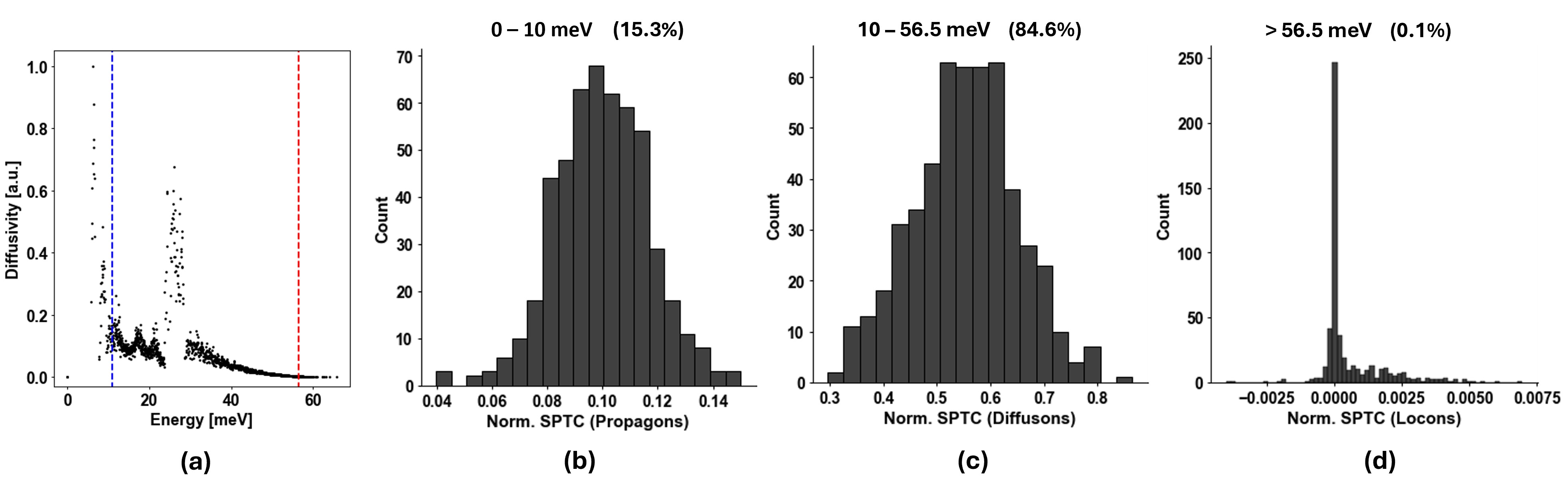}
    \caption{Vibrational mode analysis for the 512-atom a-Si model. (a) Thermal diffusivity vs. vibrational energy, with propagons (left of blue dashed line), diffusons (middle), and locons (right of red dashed line near the mobility edge). SPTC distributions from (b) propagons, (c) diffusons, and (d) locons. Insets indicate their energy ranges and percentage contributions to total SPTC. Similar plot for the 4096-atom is provided in the main text. The propagon and diffuson distributions are approximately Gaussian, with peak SPTC contributions near 10\% and 55\%, respectively. In contrast, the locon distribution is highly skewed, shwoing a narrow peak around 0.1\% and a long tail.}
    \label{Sfig_aSi512_modes}
\end{figure}

\begin{figure}[h!]
    \centering
    \includegraphics[width=\linewidth]{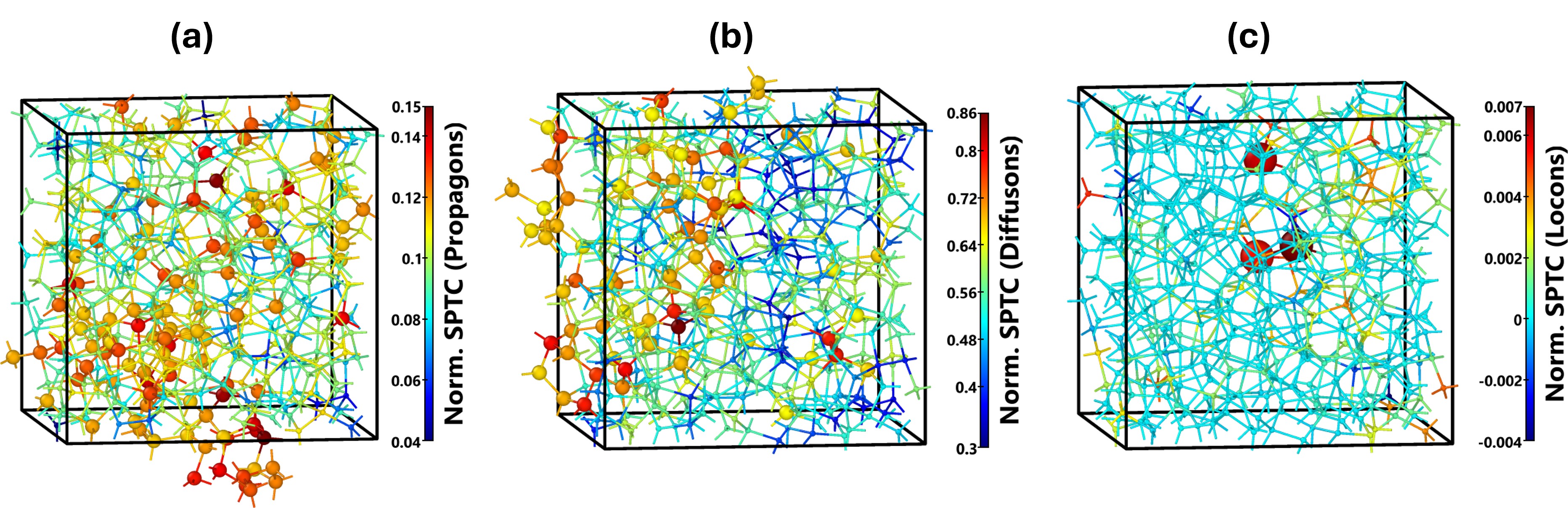}
    \caption{The contributions of (a) propagons, (b) diffusons, and (c) locons to the total SPTC are illustrated for the 512-atom model of amorphous silicon. The spatial distribution of mode-projected SPTC per atomic site is shown, with color intensity indicating the relative magnitude of each atomic site's contribution. All SPTC values are normalized by the total SPTC (i.e., without vibrational mode projection), so red regions correspond to the highest relative contributions within each mode class. Atoms contributing up to 75\% of the mode's maximum SPTC (i.e., the high-SPTC range) are highlighted with increased radius. Atomic coordinates are unwrapped from periodic boundaries to better reveal the spatial connectivity of the top 75\% contributors and their distribution throughout the simulation cell. The number of clustered or filamentary atoms (if any) in each mode is provided in Table~\ref{Stab:aSiModeFilaments}.}
    \label{fig:aSi4096_modes}
\end{figure}

\begin{figure}[h!]
    \centering
    \includegraphics[width=\linewidth]{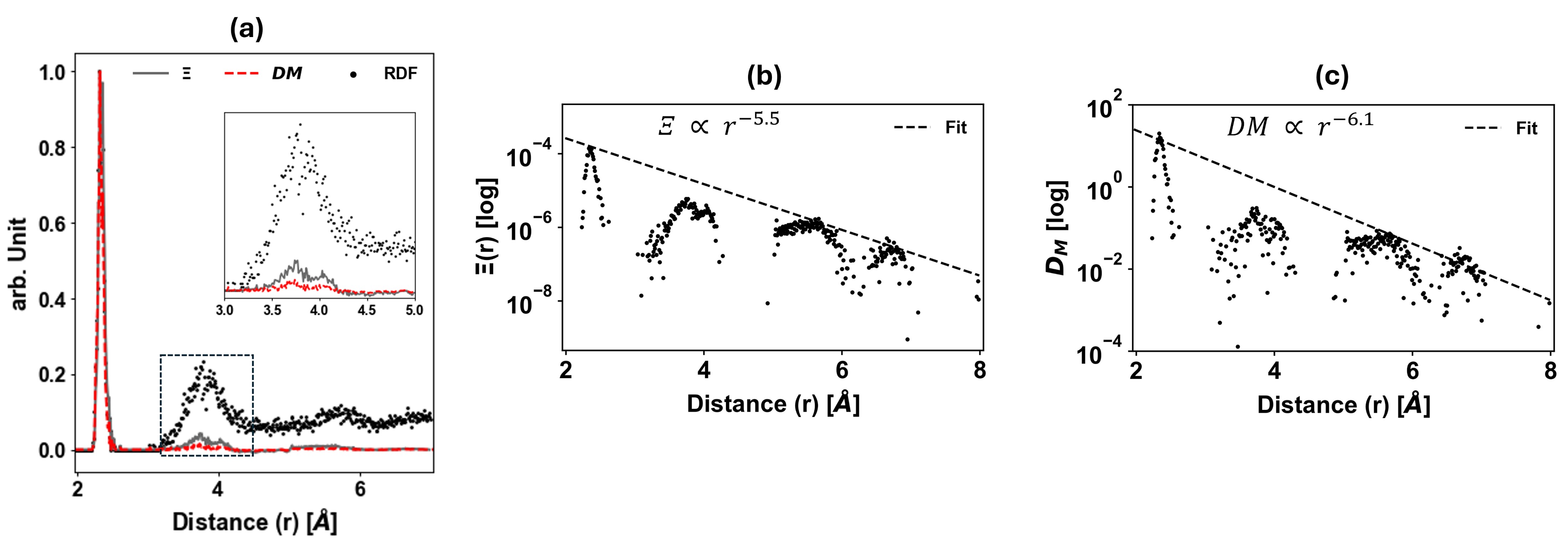}
\caption{Spatial convergence analysis of the 512-atom amorphous silicon model. (a) Cell-averaged decay of $\Xi$ and $DM$ as a function of interatomic distance $r = |x - x'|$, shown alongside the radial distribution function (RDF, black) for reference. (b) Semi-logarithmic plot (log-scale on the y-axis) of $\Xi$ illustrating exponential decay behavior, with a linear fit (dashed black line) indicating a decay rate power of 5.5. (c) Same as (b), but for $DM$, with a fitted decay power of 6.1. The decay rate for the 4096-atom model is 5.5 and 6.7 for $\Xi$ and $DM$, respectively (see main text). }
    \label{fig:Sfig_512_Xi_DM_Decay}
\end{figure}

\begin{figure}[h!]
    \centering
    \includegraphics[width=\linewidth]{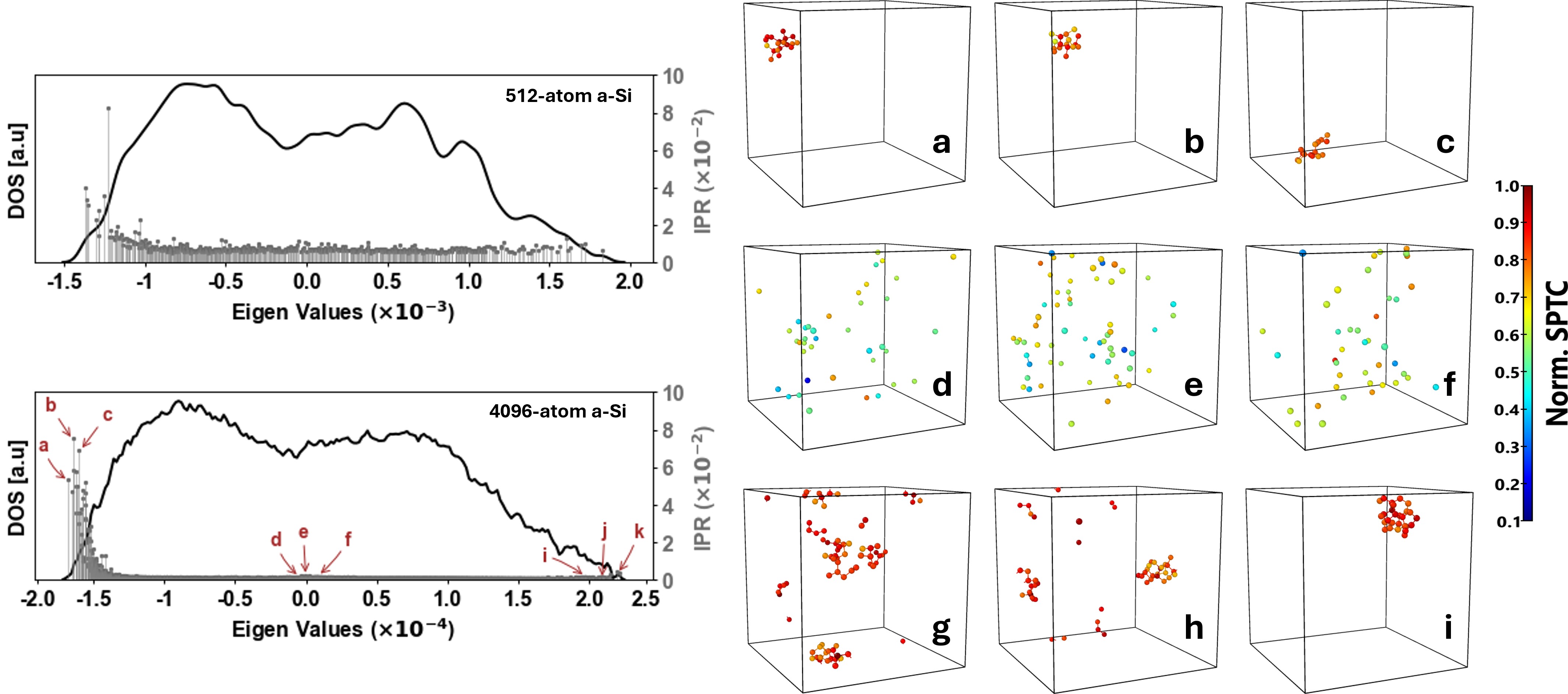}
    \caption{ The density of states (DOS; black) and inverse participation ratio (IPR; gray) of the $\Xi$ matrix, defined in Equations~\textcolor{blue}{7} and~\textcolor{blue}{15} of the main text, is shown in the top and bottom left panels for the 512- and 4096-atom amorphous silicon (a-Si) structures, respectively. Most eigenvectors of $\Xi$ are spatially extended, except for those associated with the most negative eigenvalues ($\lambda$). For the 4096-atom system, we highlight nine representative eigenvalues (\textit{a}--\textit{i}) from three distinct regions of the spectrum, and the atoms contributing to these modes are projected in the right panels as described below: (\textit{a}, \textit{b}, \textit{c}; Top) The first region, corresponding to $\lambda \ll 0$, exhibits spatially localized modes forming compact clusters (small blobs) composed of atoms with high SPTC values (see colorbar for scale).  (\textit{d}, \textit{e}, \textit{f}; Middle) The second region, where $\lambda \rightarrow 0$, features fully delocalized modes, as indicated by the DOS, involving atoms with low to intermediate SPTC values. (\textit{g}, \textit{h}, \textit{i}; Bottom) The final region, with $\lambda \gg 0$, also includes high-SPTC atoms. However, unlike the first region, these atoms generally do not form spatially connected clusters, indicating a different localization character. A similar pattern is observed for the 512-atom a-Si structure.}
    \label{fig:Sfig_aSi-Xi-spectral}
\end{figure}

\begin{figure}[t!]
    \centering
    \includegraphics[width=\linewidth]{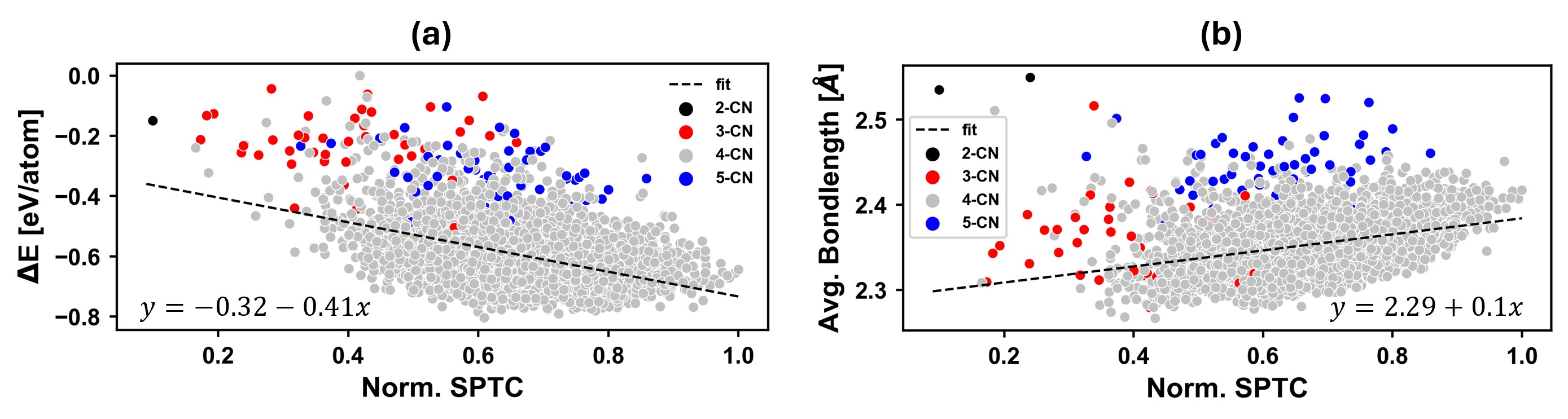}
    \caption{Thermal activity in a 4096-atom model of amorphous silicon. (a) Local atomic energy difference (\(\Delta E\)) and (b) average bond length per atom plotted against Site-Projected Thermal Conductivity (SPTC), with atoms categorized by coordination number (CN). Black dashed lines indicate linear regressions; equations are provided in the insets.}
    \label{fig:Sfig_SPTC_aSi4096}
\end{figure}

\begin{figure}[t!]
    \centering
    \includegraphics[width=\linewidth]{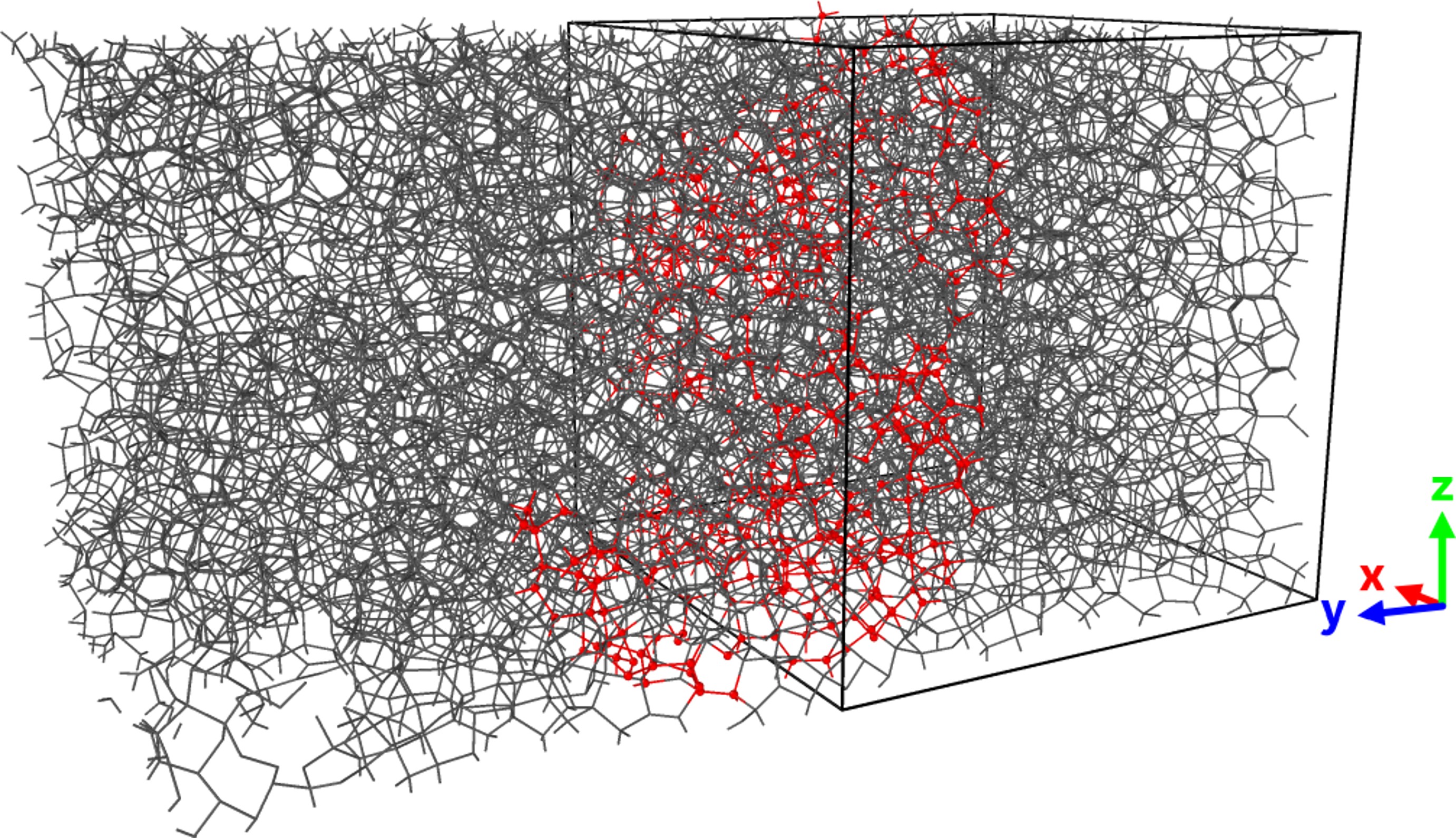}
    \caption{Heat transport network in the 4096-atom amorphous silicon model. The largest high-SPTC filament (SPTC \(>\) 75\%) is shown in red, with all other atoms rendered in gray. The supercell is outlined by a black box and repeated along the \(y\)-axis to illustrate the periodic extension of the filament structure.}
    \label{fig:Sfig_aSiHeatNetwork}
\end{figure}
